\documentclass{optica-article}

\journal{opticajournal} 

\articletype{Research Article}

\usepackage{lineno}

\usepackage{graphicx} 
\usepackage{float}
\usepackage{svg, standalone}
\usepackage{tikz}
\usetikzlibrary{arrows.meta,positioning,calc,fit}
\usepackage{mathrsfs, amsmath, gensymb}
\usepackage{siunitx}
\usepackage{layouts}

\newcommand{\be}{\begin{equation}}
\newcommand{\ee}{\end{equation}}
\newcommand{\bea}{\begin{eqnarray}}
\newcommand{\eea}{\end{eqnarray}}
\newcommand{\nn}{\nonumber}

\usepackage{subcaption}
\usepackage{ragged2e}

\begin{document}

\title{Fast projections of two-dimensional light patterns using acousto-optical deflectors}

\author{Robbert Decruyenaere\authormark{1}, Clara Tanghe\authormark{2}, Senne Van Wellen \authormark{1} and Karel~Van~Acoleyen\authormark{1,2}}

\address{\authormark{1}Department~of~Electronics~and~Information~Systems, Ghent University,~Technologiepark-Zwijnaarde~126,~9052 Ghent,~Belgium\\
\authormark{2}Department~of~Physics~and~Astronomy, Ghent University,~Krijgslaan~299,~9000~Ghent,~Belgium
}



\begin{abstract*} 
Precise and flexible control of structured light fields is essential for applications
ranging from optical trapping and quantum simulation to microscopy and materials processing. Acousto‑optical deflectors (AODs) are widely used in these settings due to their high speed, large damage threshold, and ability to generate steerable optical tweezers. Multi‑tone driving offers a powerful alternative to slow sequential scanning, enabling the projection of complex patterns with high accuracy as rapid acoustic modulation averages out inter‑spot interference. In two dimensions, however, intermodulation between tones in orthogonal AODs can reintroduce coherent artifacts.
We present a fast, feedback‑free AOD projection scheme based on an incommensurately staggered frequency lattice that intrinsically suppresses such artifacts.  
For separable two‑dimensional target patterns, our method removes the need for scanning entirely, enabling substantially faster and highly accurate projections. We further extend the approach to non‑separable images using a minimal scanning strategy that maintains rather high projection speeds. These results demonstrate that appropriately engineered multi‑tone AOD driving offers an efficient and robust route to high‑speed, high‑fidelity generation of arbitrary intensity patterns.

\end{abstract*}

\section{Introduction}

Our ability to explore the mesoscopic and microscopic world relies critically on the increasingly precise control of light fields. Optical tweezers, for instance, have found widespread applications, including the trapping of various types of particles \cite{Ashkin1970,Ni2021, Cicali2025, Fan2026}, force measurements on microscopic objects \cite{Bola2020, Bustamante2021} and targeted activation of neurons \cite{Ricci2024}. More sophisticated light fields, such as programmable tweezer arrays, optical lattices, and arbitrary continuous intensity patterns, have been further developed in areas including quantum many-body physics \cite{Schnelle2008, Henderson2009, Trypogeorgos2013, Bell2016, Barredo_2016, Browaeys2020, Ebadi2021,  Kaufman2021, Amico2021}, quantum computing \cite{Weiss2017,Henriet2020,Bluvstein_2022,bluvstein2025_qucomp}, materials science \cite{Jiao2023, Mauclair2025}, biophysics \cite{Tanaka2011, Xin2020}, and life sciences \cite{Lenton2020, Marchese2024, Reddy2005}.

The most commonly used optical sculpting devices are phase-modulating liquid-crystal spatial light modulators (LC-SLMs), amplitude-modulating digital micro-mirror devices (DMDs) and accousto-optical deflectors (AODs). LC-SLMs yield high resolution flexible light-fields, but are limited by their slow response time, up to 0.1-1~kHz \cite{Gauthier2021, Mauclair2025}. DMDs on the other hand are much faster, possessing refresh rates exceeding 20~kHz \cite{Gauthier2021, Amico2021, Mauclair2025}. A disadvantage being their low light-efficiency. They can also project arbitrary intensity patterns with high resolution. Although they are inherently binary devices, by overlapping spots from several mirrors or by time-averaging, they also allow gray-scale control. A common feature of both devices is their sensitivity to speckle caused by interference effects from shaping coherent laser light into dense patterns \cite{Akemann2024, Mauclair2025}. This speckle can be suppressed by employing polychromatic light sources \cite{Deng2017,Lee2020,Smith2021, Calzavara2023}, but this in turn introduces chromatic aberrations. In practice this means that both for LC-SLM and DMD
projections one often has to resort to camera feedback schemes for improving the
projected image towards the desired accuracy \cite{Nogrette2014, Gauthier2021, Ebadi2021, Schroff2023, Chew2024,Bruce2015, Calzavara2023}. 

Finally, AODs are one-dimensional (1D) deflector devices, where a traveling acoustic wave in the crystal acts as a (Bragg) diffraction grating for an incoming laser beam, allowing control on both the amplitude and phase profile. Compared to LC-SLMs and DMDs they have high transmission ratios and optical damage thresholds \cite{Gauthier2021}. Employing two perpendicular AODs in series enables two-dimensional (2D) control, albeit limited to separable patterns in the image plane. They are the device of choice for fast projection of single optical tweezers, used for instance for shuffling atoms around in Rydberg arrays \cite{Browaeys2020, Ebadi2021, Henriet2020, Bluvstein_2022, bluvstein2025_qucomp}. Such a tweezer (or spot in the 2D image plane) arises from driving both AODs with a single tone. The effective refresh rate in that case is set by the time required for the sound wave to pass through the beam, which for typical applications can be as small as $5~\mu\textrm{s}$. More complicated patterns can be created by the sequential projection of single spots at different locations, with the target pattern realized upon time-averaging. However, for such spot scanning scheme to be useful, the total projection period should be much smaller than the relevant time-scales of the system at hand. In the context of optical traps for instance, scanning rates that are too small can induce unwanted particle diffusion, micromotion and heating \cite{Schnelle2008, Bola2020, Bell2016, Bell2018, Ni2021, Gosar2022}.

A more attractive, faster option for generating arbitrary patterns, involves multi-tone driving of the AODs \cite{Trypogeorgos2013, Bola2020, Treptow2021, Fatemi2007, Akemann2024}. Multi-tone driving results in a superposition of spots, with the inter-spot interference —which would lead to speckle for a purely static projection— averaging out over a period set by the inverse frequency difference of the spot pairs. This allows for simple forward projection schemes that do not require feedback. Such constructions for accurate arbitrary 1D multi-toned AOD projections were presented in  \cite{Trypogeorgos2013} and \cite{Treptow2021}, with the former approach directly employing the multi-tone representation and the latter relying on a holographic parameterization of the AOD driving signal. When extending this approach to the 2D case, one must account for the intermodulation of tones in both AODs. For certain frequency combinations this can lead to \emph{coherent artifacts}, inter-spot interferences that do not average out to zero \cite{Trypogeorgos2013, Treptow2021, bluvstein2025_qucomp}, leading to speckle projection degradation. The approach of \cite{Treptow2021} works around this problem by using line-scanning, with the desired 2D pattern sequentially projected line by line.

In this paper we introduce a different scheme based on an appropriate restricted subset of AOD driving frequencies, to suppress the coherent artifacts. For separable 2D target images, $I(x,y)=I(x)I(y)$, our scheme does not require scanning, which leads to considerably faster projections for the same image accuracies. Extending our approach to non-separable patterns requires a scanning scheme, but as we will show, this still enables reasonably fast projections — faster than line-scanning schemes. 

The next section  contains the description of our experimental setup. In \autoref{sec: 2D patterns}, we consider generic periodic 2D multi-tone driving, with a particular attention to the coherent artifacts, including their experimental measurement. In \autoref{sec: incomm staggering} we then present our construction for 2D projections based on incommensurately staggered frequency lattices. We derive the trade-off between the consequential suppression of the coherent artifacts and the AOD driving period, and demonstrate this on two separable example patterns. The comparison with line-scanning protocols is reserved for \autoref{sec: comparison}. In \autoref{section: non sep}, we then show how our construction can be extended to the projection of non-separable patterns. The conclusions and outlook are presented in \autoref{sec: conclusions}. Furthermore,  In \autoref{appendix: 1D} and \autoref{appendix: 2D AOD formula}, we consider the case of 1D multi-tone driving and derive the relevant formula for our results in the main text. A version of the Gerchberg-Saxton algorithm that we use for phase-optimization is discussed in \autoref{appendix: gs}. Finally, \autoref{appendix: coherentartifact} presents more details on our measurement of the coherent artifacts.

\section{Experimental setup}\label{appendix: exp setup}

\begin{figure}[t]
\centering\includegraphics[width=10cm]{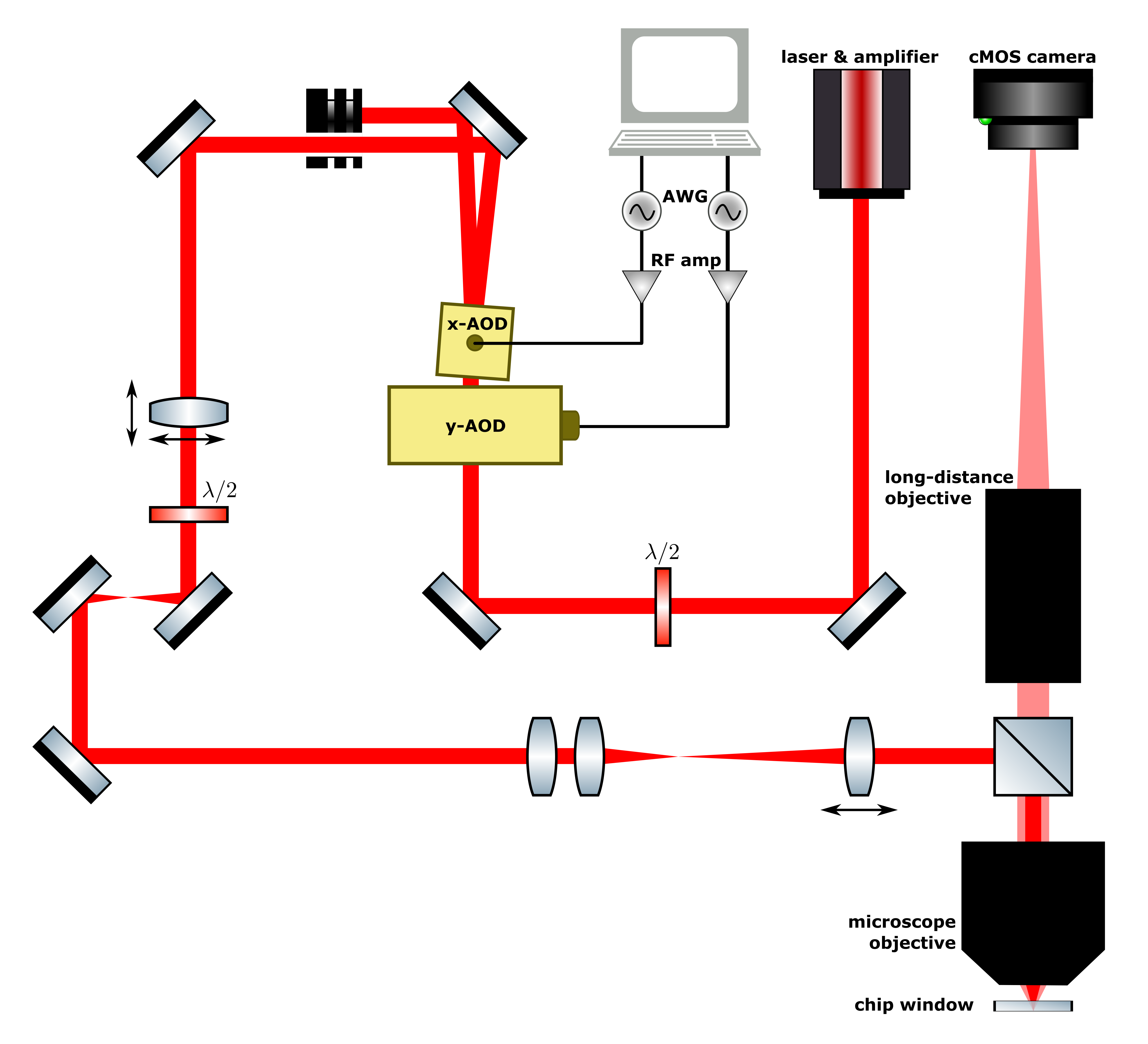}
\caption{\textbf{Optical setup.} The laser beam gets diffracted in the AODs and the (1,1) order is relayed through telescope optics and passes a coaxial illuminator which directs the light to the microscope objective (MY20X-824 - 20X Mitutoyo Plan Apochromat Objective, $0.40$ $NA$, $20.0 mm$ $WD$), focusing the light onto the desired image plane. The coaxial illuminator also allows for the light reflected of the glass plate to be directed through a long distance objective (Infinity Photo-Optical KC VideoMax lens system) which focuses the light onto the sCMOS sensor.}\label{fig: setup}
\end{figure}

The core of our experimental setup consists of a collimated laser beam which is modulated by two orthogonal AODs and then focused by a Fourier lens onto an image plane. This optical setup is designed to project arbitrary light fields on cold atoms, but we stress that all results in this paper are applicable to generic 2D AOD projections, regardless of their particular use.   

The projected light fields are generated by modulating a narrowband continuous laser beam (Toptica BoosTApro, $\lambda_0=768$ nm, FWHM = 2 nm) using two perpendicular AODs  (Gooch \& Housego, AODF 4055-4, $52.5 $MHz center frequency), see \autoref{fig: setup}.
The AODs are driven by computer-generated radio-frequency (RF) signals, converted by an arbitrary waveform generator (AWG, National Instruments Corp., 2-channel PXIe-5433, 16 bit resolution, $250 $MHz samplerate). The driving period is chosen as an integer number of AWG samples to ensure periodic waveform generation. Since the AOD bandwidth (35-70 MHz) is well below the Nyquist frequency, discretization effects are negligible and the signals are treated as continuous.
The subsequent RF amplification (Infleqtion PXIe-1300) of the acoustic signal was restricted to remain within the linear regime of the AODs, thus diffracting the incoming light in only zeroth and first order. The resulting (1,1)-order light field is focused with a Fourier lens and relay optics on the bottom side of a $400 \mu m$-thick glass plate which serves as the image plane. The reflected light is directed onto a sCMOS camera (Oxford Instruments Andor Zyla 4.2). 
Note that this configuration allows us to image the AOD projections as shown in this paper. The cold atoms in our experiment are actually located $\sim 100\mu m$ below the chip window. To project light fields on the cold atoms, the focus is changed such that the image plane coincides with the atoms, see \cite{Tanghe2025} for a description of the complete experiment. 

\begin{figure}[t]
    \begin{subfigure}{0.33\textwidth}
    \centering
    \includegraphics[width=\textwidth]{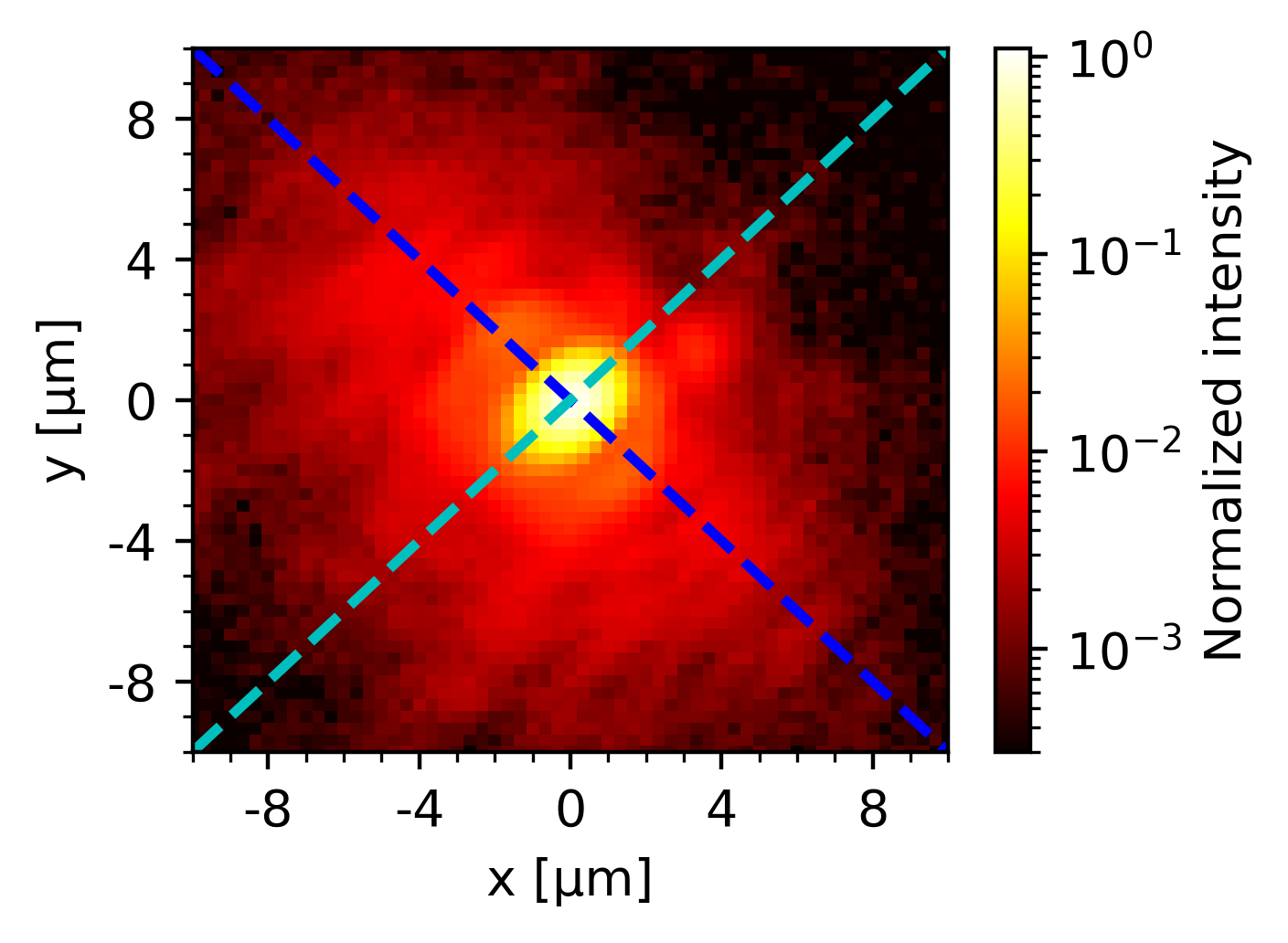}
    \caption{}
      \end{subfigure}
  \hfill
  \begin{subfigure}{0.32\textwidth}
    \centering
    \includegraphics[width=\textwidth]{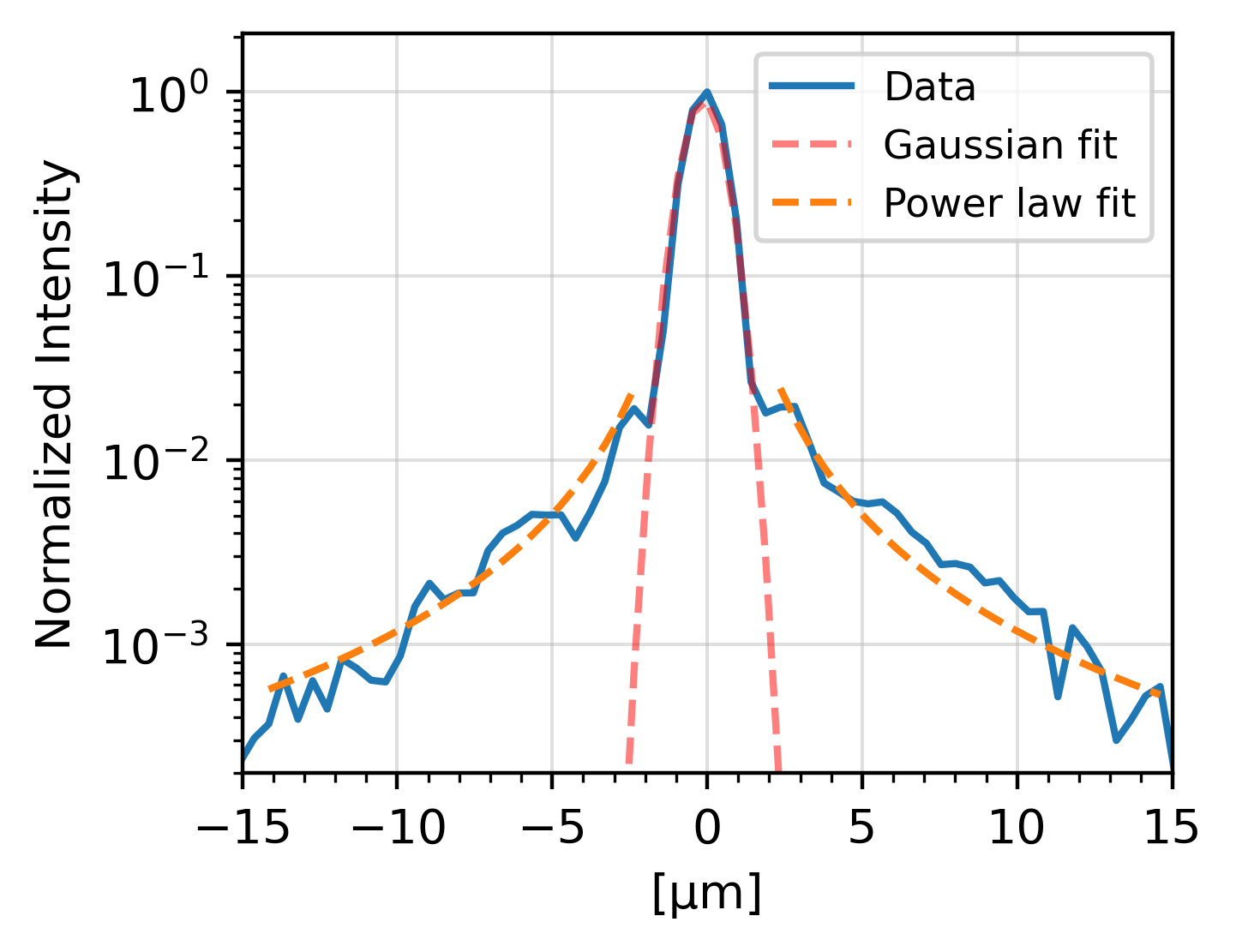}
    \caption{}
      \end{subfigure}
  \hfill
  \begin{subfigure}{0.32\textwidth}
    \centering
    \includegraphics[width=\textwidth]{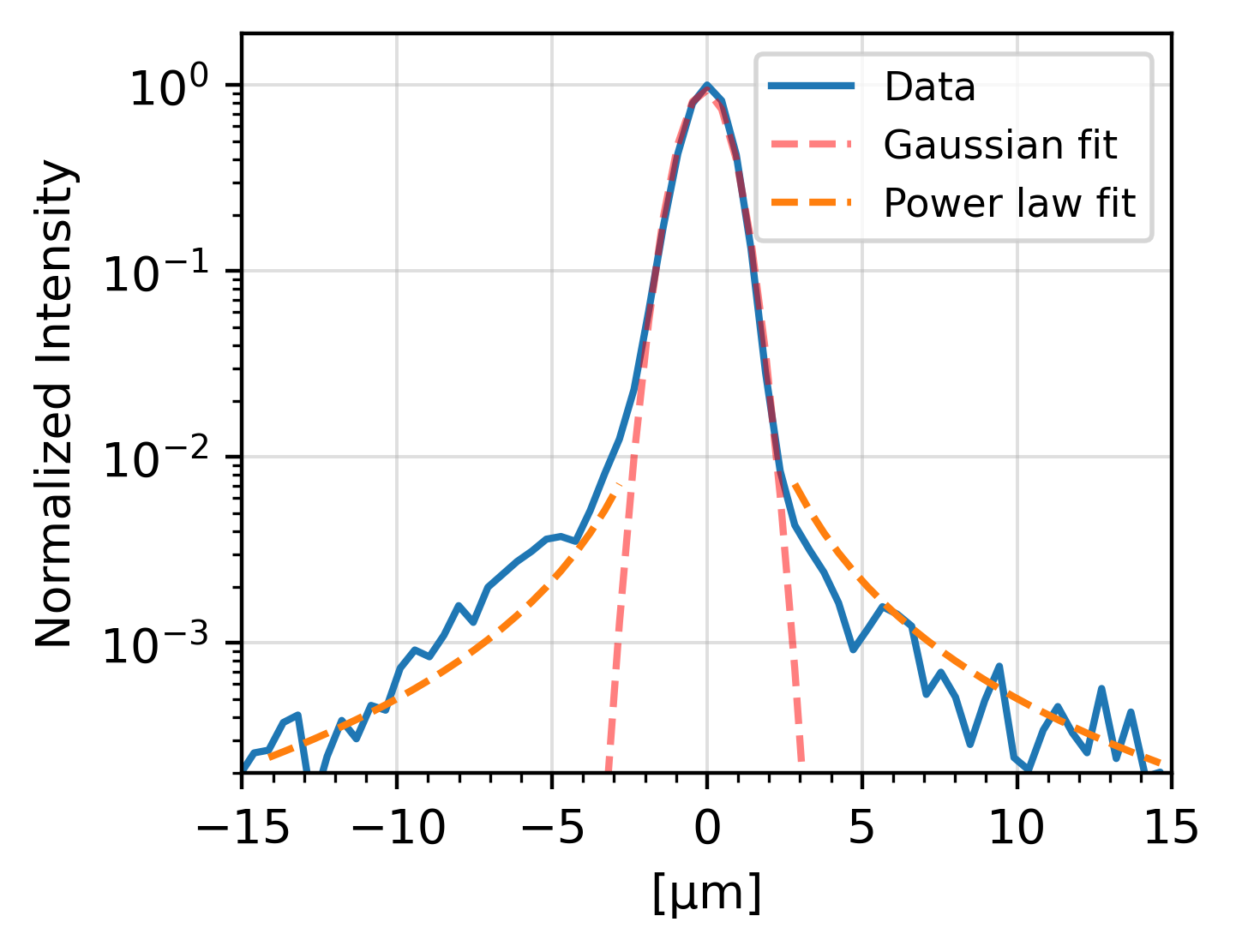}
    \caption{}
      \end{subfigure}
     
  \caption{\textbf{Single spot in the focal plane} (a) Normalized 2D intensity map. (b) Cross section, in log-scale, along blue dashed line marked in (a). The central region is fitted with a gaussian function $I(x)\sim \exp(-2x^2/w^2)$ with  waist $ w = 1.18\pm0.05 \mu m$. The outer region is fitted with a power law $I(x)\sim |x|^{-\kappa}$ with $\kappa = 2.10\pm0.10$. 
  (c) Cross-section along cyan dashed line. The central and outer regions are respectively fitted with a gaussian function $I(x)\sim \exp(-2x^2/w^2)\text{ with }w = 1.52\pm 0.04~\mu m$ and a power law $\ I(x)\sim |x|^{-\kappa} \text{ with }\kappa = 2.10\pm 0.15 $. }  
  \label{fig: spot}
\end{figure}

In \autoref{fig: spot} we show the resulting image for a single spot, obtained by driving both AODs with a single tone. This effectively measures the amplitude of the system's Point Spread Function (PSF) as the intensity distribution is directly proportional to the squared PSF amplitude, $I(x,y)\sim |PSF(x,y)|^2$. We find an approximate Gaussian profile near the center, dressed by power-law $\sim 1/r^2$ tails at large distances $r$ from the center. Such fat tails are generally expected for clipped Gaussian beams \cite{Fourier_Goodman1969}. In this case, the finite-sized AODs are providing the main clipping and as we show in \autoref{sec: 2D patterns}, the ensuing tails lead to sizable interference effects for spots that are already well separated. 

\section{Generic 2D multi-tone driving}\label{sec: 2D patterns}
We consider multi-tone driving for both the $x$- and $y$- axis AOD, 
\begin{align}
s_j(t) = \sum_k a^j_k \sin(2\pi f^j_k t + \alpha^j_k)\quad\quad \textrm{with}\,\, j=x,y\,\, \label{eq:2Dmultitone},
\end{align}
with $s_j(t)$ the RF-signals that are transmitted through the respective AODs (see \autoref{appendix: exp setup}). 
As shown in \autoref{appendix: 2D AOD formula}, in the linear regime, the light field in the image plane resulting from the first diffraction orders in both AODs, reads:
\be U(x, y, t) = \sum_{k,p}A^x_kA^y_p PSF(x-x_k, y-y_p) e^{2\pi i(f^x_k+f^y_p)t}\,,\ee
with
\be A^j_m = c^j_m a^j_m e^{i\alpha^j_m}\,\, \quad\text{and}\quad x_k = C f^x_k\,, y_p=Cf^y_p\,.\ee 
The pre-factors $c^j_k$ parameterize the frequency-dependent diffraction efficiencies of both AOD devices \cite{NAntonov2019}. These diffraction efficiency maps $\{c_k^x, c_p^y\}$ for both x- and y-axis AOD are easily determined from measuring the intensity of single-tone drives at different frequencies, see e.g. \cite{Treptow2021}. The constant $C$ captures the linear relation between the AOD frequencies and the corresponding translation of the PSF in the image plane. For our system, $C=3.864$~$\mu$m/MHz. Crucially, the generated light field (omitting the ultra-fast $\sim 300$~THz dependence of the 768~nm light) oscillates in time, with the field of a spot located at ($x_k,y_p$), oscillating with frequency $f^x_k + f^y_p$.

The actual light intensity $I(x,y,t)=U(x,y,t)\bar U(x,y,t)$ in the image plane then reads:
\be I(x,y,t) = \sum_{k,l,p,q} A^x_k\bar{A}^x_l A^y_p\bar{A}^y_q PSF(x-x_k,y-y_p) \overline{PSF}(x-x_l,y-y_q) e^{2 \pi i  \left(f^x_k-f^x_l+f^y_p-f^y_q \right)t} \,.\label{eq: Int2D}\ee

In this sum we can distinguish the single-spot contributions, $(k,p)=(l,q)$, from the other terms that are the interference fringes formed by two partially separated coherent beams. Furthermore, we restrict to periodic driving with a common period for both signals $s_j(t)=s_j(t+\tau)$, or equivalently $f^j_k=n^j_k/\tau$ with $n^j_k\in \mathbb{N}$. This intensity can then also be decomposed as a sum of a time-independent part, the average intensity $I_{avg}(x,y)$, and a time-dependent part $\tilde{I}(x, y,t)$, which averages to zero over one period:
\begin{align}
    I(x,y,t) = I_{avg}(x,y) +\tilde{I}(x,y,t) \,.
\end{align}
Importantly, in contrast to the 1D case (see \autoref{appendix: 1D}), we get coherent artifacts \cite{Trypogeorgos2013,Treptow2021}, i.e. extra static interference fringes $I_C(x,y)$ that contribute to the average intensity, in addition to the \emph {incoherent} single-spot terms $I_{IC}(x,y)$:
\be
    I_{avg}(x,y) =\frac{1}{\tau}\int^{t_0+\tau}_{t_0}\! dt\, I(x,y,t)= I_{IC}(x,y) + I_C(x,y)\,.\label{eq: Iavg2D}
\ee    
Let us first focus on the incoherent part: \be 
    I_{IC}(x,y) = \sum_{k,p} |A^x_k|^2|A^y_p|^2|PSF(x-x_k,y-y_p)|^2 \label{eq: IC2D}\,,\ee
and assume that we can actually neglect the coherent artifacts. In that case the average intensity, $I_{avg}\approx I_{IC}$, is composed of individual spots, forming a 2D lattice with spot positions $(x_k, y_p)$. For the case of equidistant frequencies, separated by $\Delta f_{min}$, the spots are separated by the \emph{spot pitch} $a$, i.e. the distance between two neighboring spots: 
\be a = C \Delta f_{min} = C/\tau \,.\label{eq: spot pitch}\ee 
As such we already get a tension between the spatial and temporal resolution of the projection. But notice that the final spatial resolution is also limited by the intrinsic optical resolution. In particular, once the spot pitch is of the order of the spot waist, $a\sim w $, a further decrease of the spot pitch will not considerably improve the spatial resolution anymore. 

Importantly, as $I_{IC}$ only depends on the AOD amplitudes and PSF amplitude, it lends itself to straightforward optimization of the AOD amplitudes for a given target pattern, via least squares minimization of $\iint\! dxdy\, (I_{IC}(x,y)-I_{tar}(x,y))^2$. Note here that, inherent to the 2D AOD setup, the control on the amplitude of the different spots is limited to a separable form $\sim |A^x_k|^2|A^y_p|^2$. Therefore one can only expect $I_{IC}$ to yield a good approximation to (almost) separable target patterns, $I_{tar}(x,y)\approx I^x_{tar}(x) I^y_{tar}(y)$. 

Furthermore, the phases $\alpha^j_k$, which do not enter $I_{IC}(x,y)$, can be chosen to reduce the amplitude variation of the driving signals $s_j(t)$. This maximizes the power efficiency, pushing the signals towards the linear AOD regime and thereby avoid nonlinear distortion \cite{Sun2024}. 
Several techniques have been proposed for this purpose \cite{yang2015, Sun2024}, in \autoref{appendix: gs} we show a variation of the Gerchberg-Saxton algorithm used in this work. Note that, in the case of 1D multi-tone driving, the approach of \cite{Treptow2021} employs a different holographic local parameterization of the periodic signal $s(t)=s(t+\tau)$ to achieve the same goal.

Of course, a projection scheme based on the incoherent part in the averaged image \eqref{eq: IC2D}, relies on the coherent part being small, which will not be the case for generic choices of the AOD driving frequencies. In the rest of this section we further investigate the precise form and magnitude of this coherent artifact $I_C$. This will motivate our construction of an AOD driving scheme in the next section, that suppresses the effect of the coherent terms, and therefore allows optimization on the incoherent part \eqref{eq: IC2D} as explained above.

This time-independent coherent part $I_C(x,y)$ consists of all the terms in Eq. \eqref{eq: Int2D} for which the two interfering spots with frequency pairs $(f^x_k,f^y_p)$, $(f^x_k, f^y_q)$ lie along a diagonal,  \be f^x_k-f^x_l=f^y_q-f^y_p\,, \label{eq: condfreq} \ee which corresponds to a diagonal $\Delta x=-\Delta y$ in the image plane. The explicit expression reads (see also \cite{Treptow2021}):
\begin{align}
    I_C(x,y) = \sum_{\substack{k\neq l,\ p\neq q \\ f^x_k-f^x_l=f^y_q-f^y_p}}|A^x_kA^x_lA^y_pA^y_q|O_{klpq}(x, y) \cos \left(\alpha^\text{tot}_{kplq}+\theta_{klpq}(x,y)\right)\,, \label{eq: artifact intensity}\end{align}
    with:
\begin{align} O_{klpq}(x, y) &= \left|PSF(x-x_k,y-y_p)PSF(x-x_l,y-y_q)\right|\,,\nn\\
     \alpha^\text{tot}_{kpl q} &=\, \alpha^x_k+\alpha^y_p-\alpha^x_l-\alpha^y_q\,, \label{eq: phase2D}\\
     \theta_{klpq}(x,y) &= \arg \{PSF(x-x_k,y-y_p)\overline{PSF}(x-x_l,y-y_q)\}\,.\nn 
\end{align}

\begin{figure}[t]
    \centering
    \begin{subfigure}[b]{0.32\textwidth}
        \includegraphics[width=\linewidth]{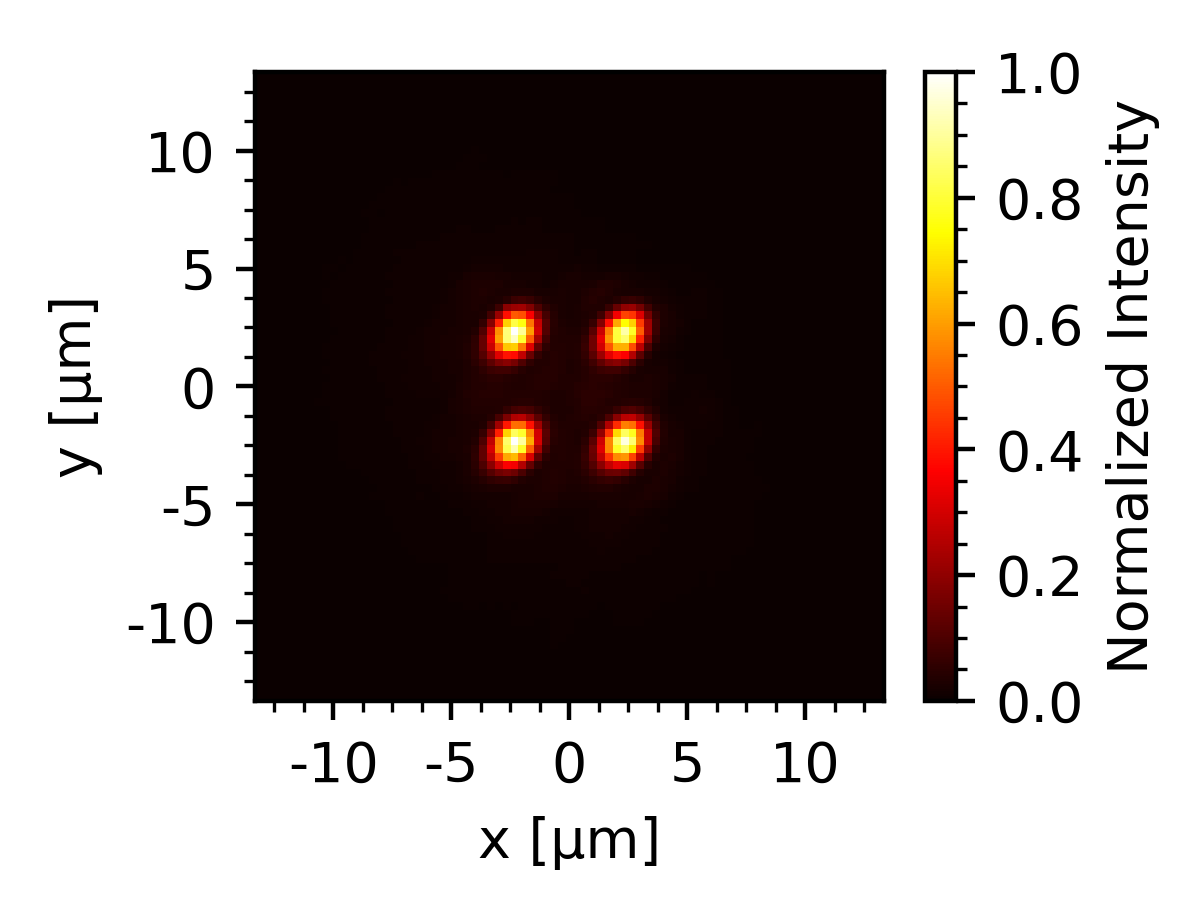}
        \caption{$I_{IC}(x,y)$}
        \label{fig:a}
    \end{subfigure}\hfill
    \begin{subfigure}{0.32\textwidth}
        \includegraphics[width=\linewidth]{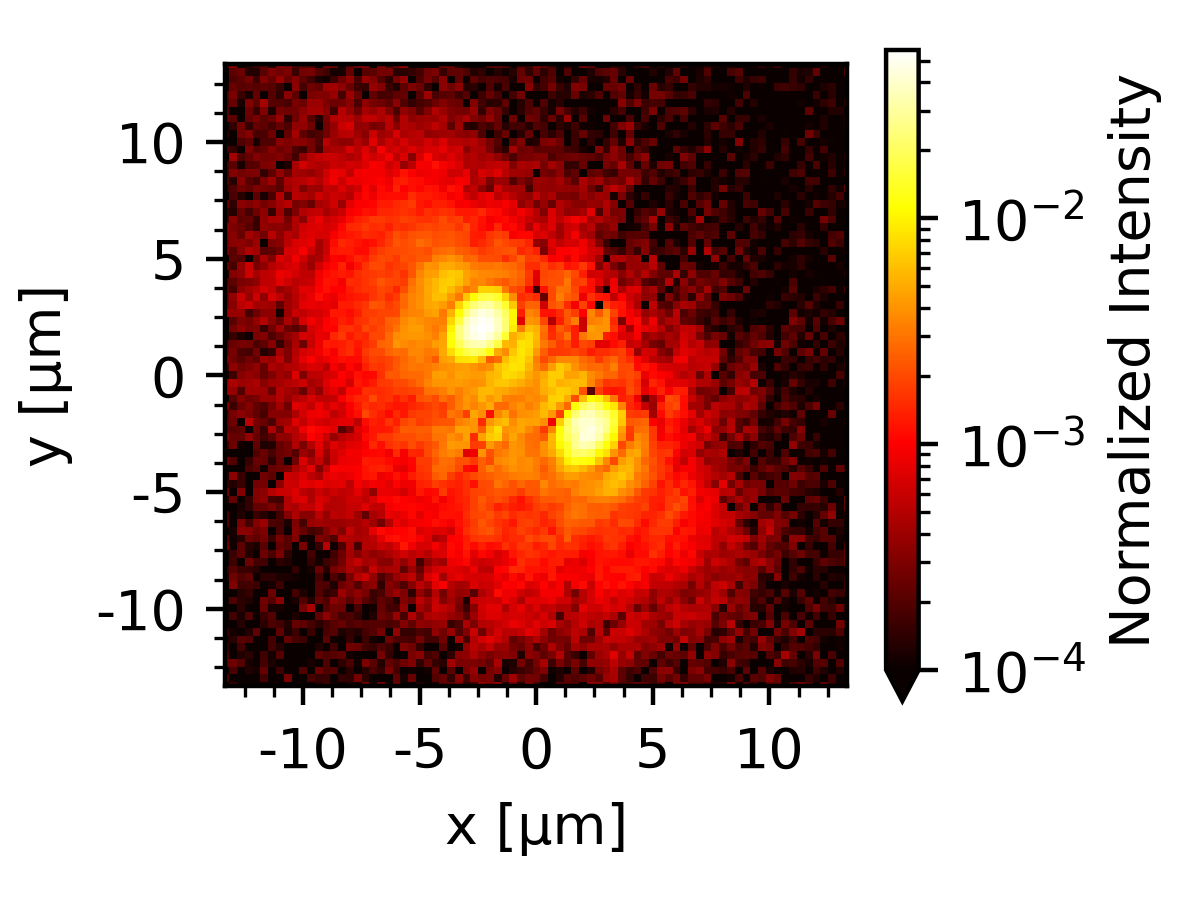}
        \caption{$O(x,y ; W\!=4.36~\mu m)$}
        \label{fig:b}
    \end{subfigure}\hfill
    \begin{subfigure}{0.32\textwidth}
        \includegraphics[width=\linewidth]{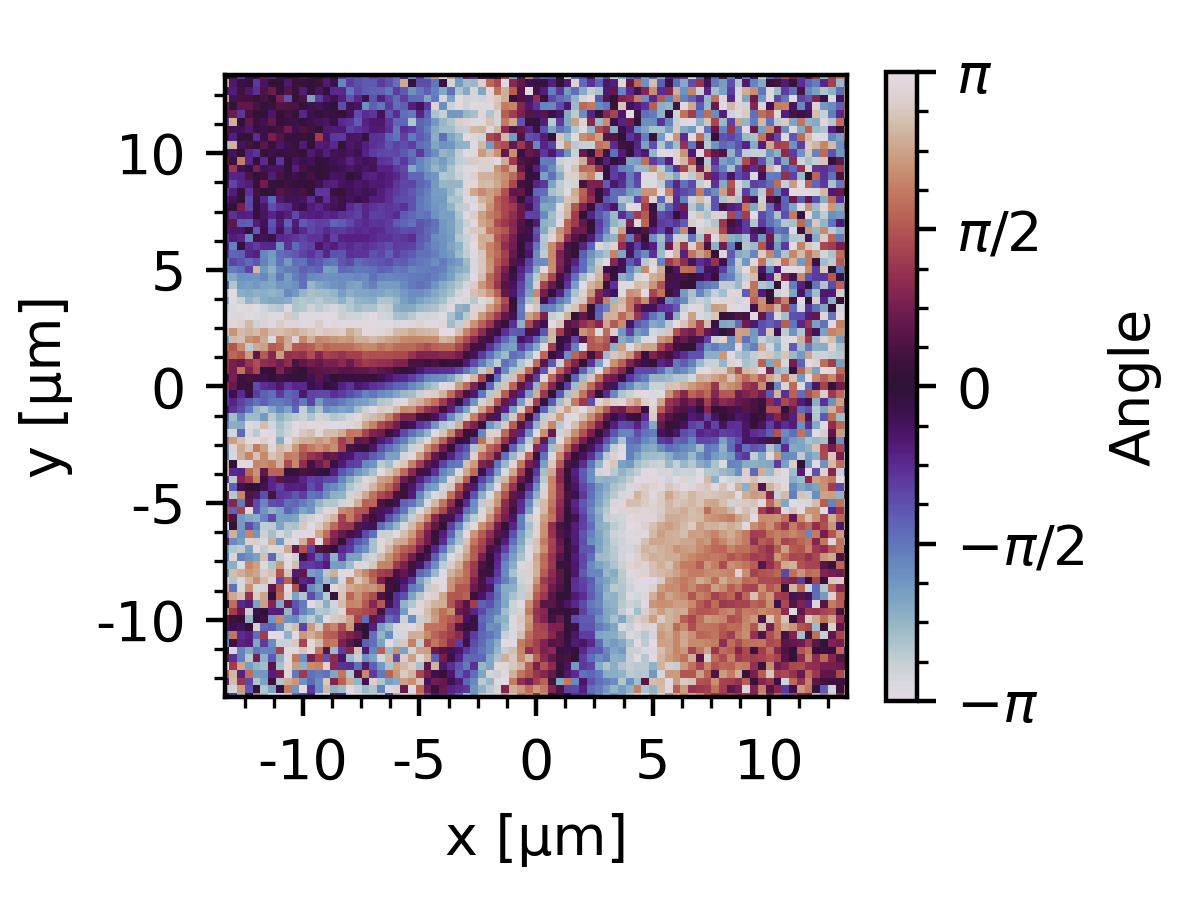}
        \caption{$\theta(x,y; W\!=4.36~\mu m)$}
        \label{fig:c}
    \end{subfigure}
    
    \caption{\textbf{Spot interference measurement.} For $W=4.36~\mu\textrm{m}$, (a) the inferred incoherent term, (b) coherent amplitude and (c) coherent phase, from fitting Eq. \eqref{eq: CAmeasurement} to 10 images that result from AOD driving with frequencies $f^{x/y}_0=f_C$, $f^{x/y}_1=f_C \pm \Delta f=f_C\pm W/C$, where $f_c=52.5$~MHz, $\Delta f=1.2~$MHz and with amplitudes \eqref{eq: AmpCA}, where $\Delta \phi=n\times 2\pi/10$, $n=0,\ldots,9$. Intensities are normalized to the center intensity of a single spot.} 
    \label{fig: 2spotInterference12}
\end{figure}

The magnitude of two interfering spots on a diagonal is essentially determined by the overlap between their respective PSFs, which in turn depends on their relative coordinate distance, $\Delta x=-\Delta y=W$. To measure the magnitude,
\be O(x,y;W)= \left|PSF(x-W/2,y+W/2)PSF(x+W/2,y-W/2)\right|\,, \ee
and phase
\be \theta(x,y;W)=\arg\{PSF(x-W/2,y+W/2)PSF(x+W/2,y-W/2)\}\,,\ee
of this overlap, we drive both AODs with two frequencies, $\{f_0^x, f_1^x\}$ and $\{f_0^y,f_1^y\}$, where $f_1^x=f_0^x+W/C$ and $f_1^y=f_0^y-W/C$. This generates four spots, of which the two spots centered at $(x_0,y_0)$ and $(x_1,y_1)=(x_0+W,y_0-W)$ will interfere in the averaged image, as they lie along a $\Delta x=-\Delta y$ diagonal. In particular, if we take 
\be A^x_0=A^x_1=A^y_0=1\,,\quad A^y_1=\exp{(i\Delta\phi)}\label{eq: AmpCA} \,,\ee the averaged intensity — under a translation in the image plane — reads (see Eqs. \eqref{eq: Iavg2D}- \eqref{eq: phase2D}): 

\be I_{avg}(x,y)=I_{IC}(x,y)+2\, O(x,y;W)\cos\left(\theta(x,y;W)+\Delta \phi\right)\,.\label{eq: CAmeasurement} \ee
The measured intensity profiles are fitted to this expression for different values of $\Delta \phi$. We can then determine the incoherent part (first term) and both the magnitude and phase of the PSF overlap (second term), see \autoref{appendix: coherentartifact} for an example fit.

\begin{figure}[t]
    \centering
    \includegraphics[width=0.5\textwidth]{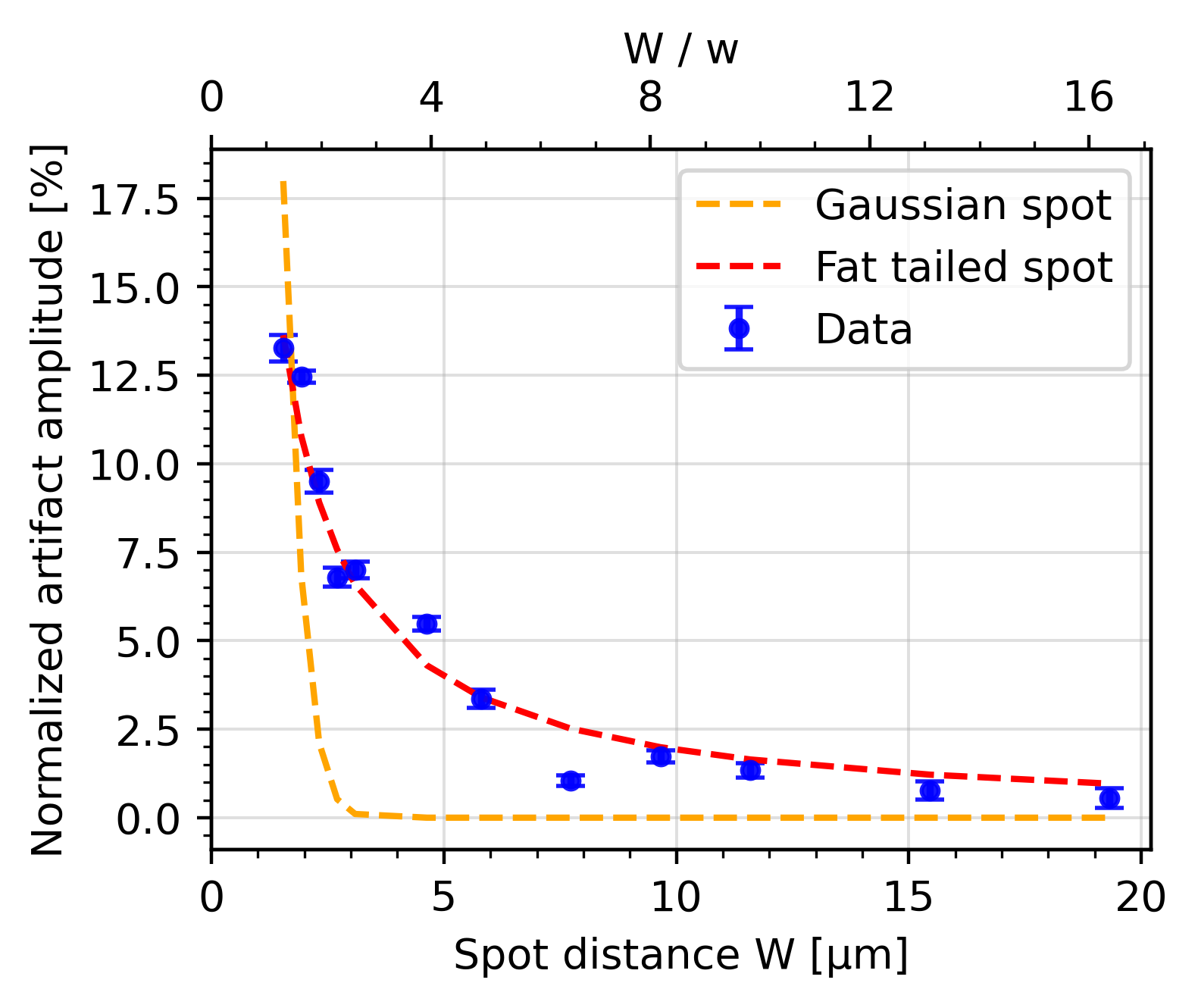}
    \caption{\textbf {Spot interference amplitude. } 
    Data points show the measured maximum of the overlap function $O(x,y;W)$, which is found at the center of the upper left spot for all considered spot distances $W$, displayed both in units [µm] and in units of the (gaussian) spot waist $w$. The orange dashed line shows the amplitude expectation from the gaussian center of the PSF, imaged in \autoref{fig: spot}, $max_{x,y}\{O(x,y;W)=\exp{(-W^2/2w^2)},\  w=\!1.18~\mu$m. 
    The red line shows the expectation from the fat tails of the PSF, $max_{x,y}\{O(x,y;W)\}\propto W^{-|\kappa|/2},\ \kappa=2.1$.} 
    \label{fig: 2spotInterference ampl_vs_dist}
\end{figure}

\autoref{fig: 2spotInterference12} displays the results for coordinate distance $W=4.36~\mu\textrm{m}$. In \autoref{fig:a}, one can observe the four spots in the incoherent part of the image $I_{IC}(x,y)$, while \autoref{fig:b} shows the artifact amplitude $O(x,y)$. Note that this amplitude is peaked at the locations of the two interfering spots. This is due to the fat power-law tails of the spot profiles, see \autoref{fig: spot} in \autoref{appendix: exp setup}. For strictly Gaussian spots, one can verify that the artifact amplitude should display a single peak right at the center between the two spot positions. Finally, the artifact phase $\theta(x,y)$ in \autoref{fig:c} shows the highly oscillatory behavior of the spot interference.  

In \autoref{fig: 2spotInterference ampl_vs_dist}, we display the measured artifact peak amplitude as a function of the inter-spot coordinate distance $W$, which matches well with the power-law prediction based on the measured single spot PSF of \autoref{fig: spot}. We also show the analogue prediction that only considers the central Gaussian part of the spot profile. This shows, as expected, that the coherent artifacts for clipped beams with fat tails — inherent to AOD steering — decay much slower than in the strictly Gaussian case. For instance, for $W=4.36~\mu\textrm{m}$, which is well above the diagonal spot waist $w=1.18~\mu\textrm{m}$, the coherent artifact still leads to an image degradation of more than $5\%$, whereas this would have been $1.2\cdot10^{-4}\%$ for a strictly Gaussian beam.

Our results explicitly show that coherent artifacts provide a significant contribution to the final intensity for generic 2D AOD projections. They considerably degrade the fidelity of dense projections, with spot-pitches of the order of the beam-waist ($a\sim w$) and still have a measurable impact for sparse projections ($a>w$) such as optical tweezer arrays. 

One way to reduce the coherent artifacts is by optimizing the optical setup such that the PSF has slimmer or no tails. This could be achieved for instance, by working with a smaller input beam. But note that this would come at the cost of a larger spot size and/or smaller field of view. In any case, coherent artifacts will be present for any setup, especially in dense projections. 

Although — barring other image aberrations — the explicit expression of the coherent term in Eq. \eqref{eq: Iavg2D} is known, in practice it is difficult to use it for improvement schemes of AOD projections. However, as already noted in \cite{Trypogeorgos2013,bluvstein2025_qucomp}, one can avoid the coherent artifacts altogether, by restricting to incommensurate frequencies that do not fulfill the diagonal condition \eqref{eq: condfreq}. In the following, we present a concrete scheme based on this approach for periodically driven AOD projections.

\begin{figure}[t]
     \centering
     \begin{subfigure}[b]{0.45\textwidth}
         \centering
         \includegraphics[width=\textwidth]{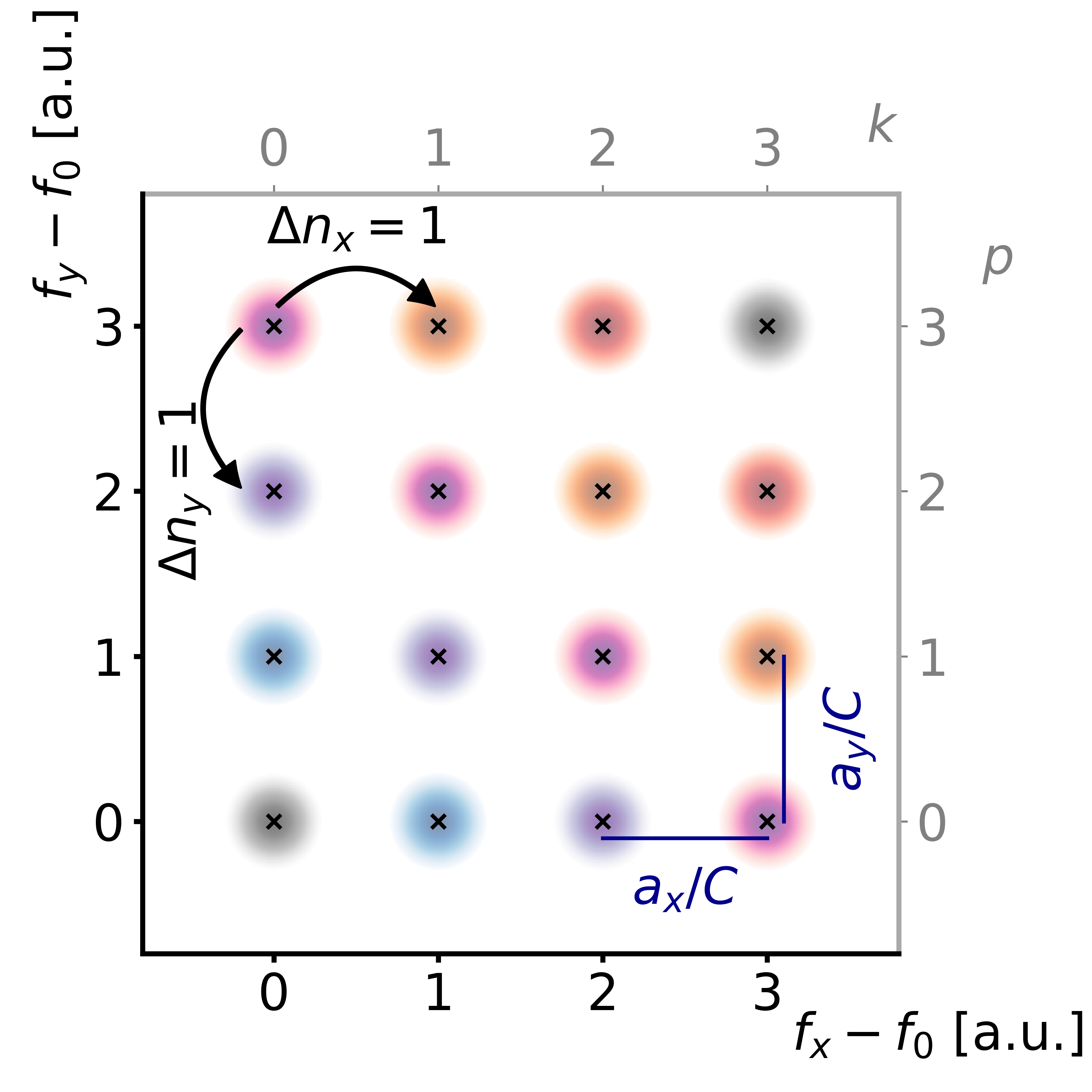}
         \caption{}
         \label{fig: Delta n^x = 1, Delta n^y=1}
     \end{subfigure}
     \hfill
     \begin{subfigure}[b]{0.45\textwidth}
         \centering
         \includegraphics[width=\textwidth]{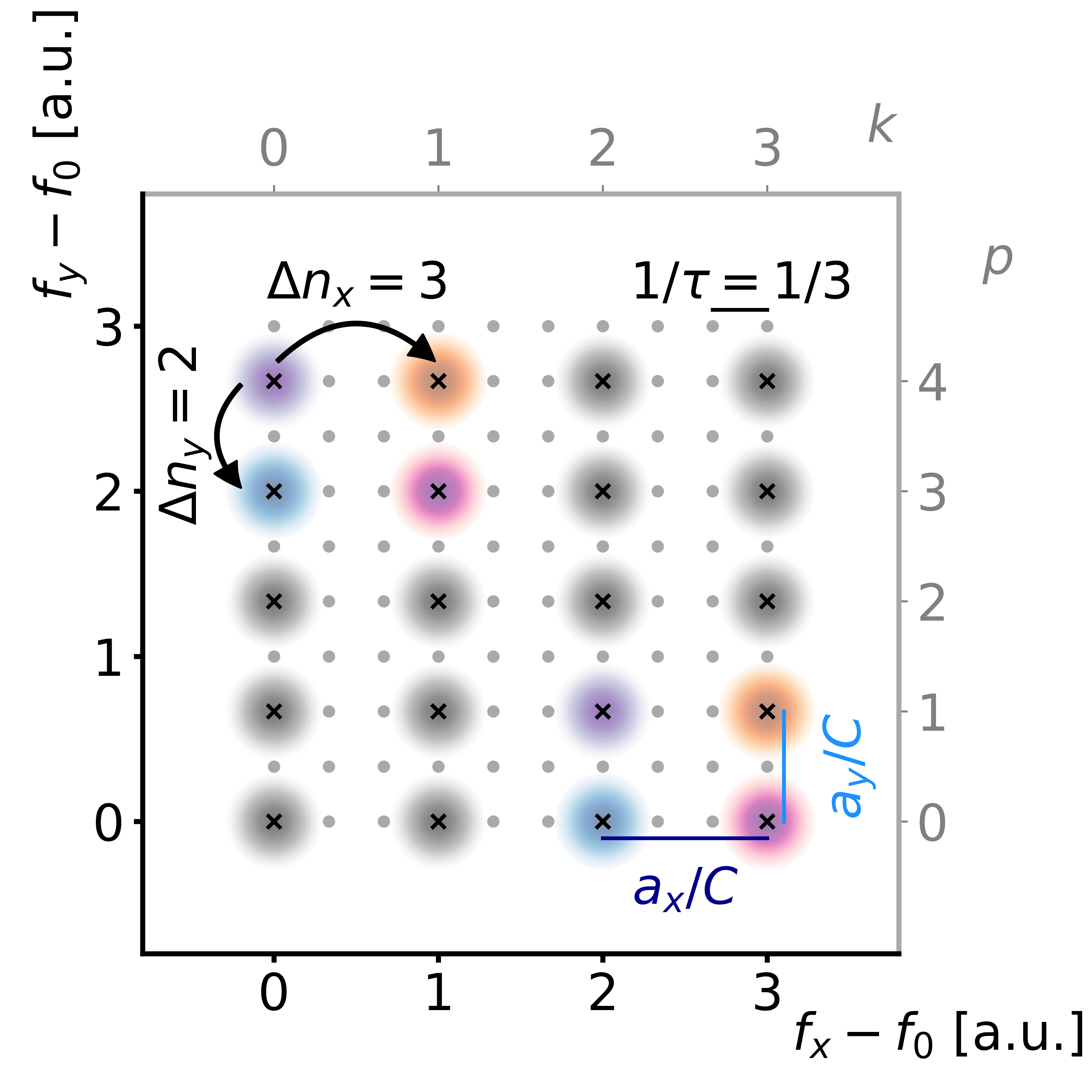}
         \caption{}
         \label{fig: Delta n^x = 3, Delta n^y=2}
     \end{subfigure}
        \caption{\textbf{Incommensurate staggering scheme.} Projecting a grid with spot pitch $a=a_x$ is achieved by periodically driving AODs along the $x$ and $y$-axis with a set of frequencies, related to the target spot positions in the image plane via $(x,y)=C(f^x, f^y)$. The frequency scale on the main axes is identical for the two figures. The numbers on the (gray) secondary axes display the lattice coordinates of the spots $(k,p)$, see Eq. \eqref{eq: staggered}. Color-values indicate interfering spots that lie on a common diagonal, $\Delta n^x \Delta k  = - \Delta n^y \Delta p,$ except for gray which indicates spots that do not share such a diagonal. (a) Regular lattice, $\Delta n_x=\Delta n_y=1$, with the nearest spots on a common interfering diagonal having lattice coordinates $(k,p)$ and $(k+1,p-1)$. (b) Incommensurate staggered lattice, $\Delta n_x=\Delta n_y+1=3$. The driving period $\tau$ is increased by a factor of 3, to keep the effective spot pitch $a_x=C\Delta f^x=C \Delta n_x/\tau$ in the $x$-direction constant. This reduces the spot pitch $a_y=C \Delta f^y= \Delta n_y/\tau$ in the $y$-direction by a factor $2/3$. On this staggered lattice, the nearest spots on a common diagonal have coordinates $(k,p)$ and $(k+2,p-3)$. As their distance in the image plane is larger with respect to the non-staggered case, their interference is suppressed. 
        Note that this illustrates a sparse projection. In case of a dense projection, there would be significant overlap between neighboring spots. }  
    \label{fig: scheme}
\end{figure}

\section{Incommensurate staggering}\label{sec: incomm staggering}
Our scheme for suppressing coherent artifacts is illustrated in \autoref{fig: scheme}. Assume we want to project an image with a certain spot pitch $a$. \autoref{fig: Delta n^x = 1, Delta n^y=1} shows the situation for a regular lattice of available frequencies with a certain periodic drive  — or equivalently, available spot positions — where all spots on a common $\Delta x=-\Delta y$ diagonal will interfere in the averaged image. \autoref{fig: Delta n^x = 3, Delta n^y=2} shows how one can suppress the coherent artifacts, without increasing the spot pitch, by driving with a larger period. This refines the lattice of available frequencies (spot positions). Restriction to a particular incommensurate staggered sublattice of this refined lattice, with asymmetric lattice spacings $a_x\neq a_y \leq a$, then enlarges the distance between spots on a common diagonal, which indeed reduces their interference amplitude, see \autoref{fig: 2spotInterference ampl_vs_dist}.
To formalize this construction, we introduce two parameters $\Delta n_x, \Delta n_y\in \mathbb{N^+}$ that define the staggering on the refined lattice. For the frequencies on the staggered sublattice we can then write: 
\be f^x_k=f_0+k \frac{\Delta n_x}{\tau}\,,\quad\quad f^y_p=f_0+p\frac{\Delta n_y}{\tau}\,,\quad\quad\quad k,p\in \mathbb{N}\,,\label{eq: staggered} \ee with $f_0\,$ a multiple of $1/\tau$. Two spots on this staggered lattice, labeled by $(k,p)$ and $(k +\Delta k,p +\Delta p)$ lie on a common $\Delta x=-\Delta y$ diagonal,
if  $\Delta n_x \Delta k=-\Delta n_y \Delta p=m$. In (refined) lattice units, this nearest distance $m$ between two such spots is found as the smallest common multiple of $\Delta n_x$ and $\Delta n_y$. Without loss of generality, we take $\Delta n_y<\Delta n_x$. In that case, for a given $\Delta n_x$, the smallest common multiple $m$ is maximal for the \emph{incommensurate staggering}: \be \Delta n_y=\Delta n_x-1\,. \label{eq: incommen}\ee In the following we will always take $\Delta n_y$ according to this condition. The corresponding smallest common multiple $m=\Delta n_x (\Delta n_x-1)$ then translates to the frequency distance $\Delta f = m/\tau$ or the coordinate distance $W_{min}=m C/\tau$ for the nearest interfering spots in the image plane. In terms of the effective spot pitch $a=a_x=C\Delta n_x/\tau$, the latter relation reads:
\be W_{min}= a\left(\Delta n_x-1\right)=a\left( \frac{a \tau}{C}-1\right)\,.\label{eq: Wmin}\ee
This makes the trade-off explicit between the AOD period $\tau$ and the suppression of coherent artifacts on the incommensurate lattice \eqref{eq: incommen}. Given a certain desired spot pitch $a$, an increase of the staggering parameter $\Delta n_x$ increases the minimal distance $W_{min}$ between interfering spots, at the price of a larger driving period $\tau\propto \Delta n_x$.

\begin{figure}[t]
    \centering
    \includegraphics[width=\textwidth]{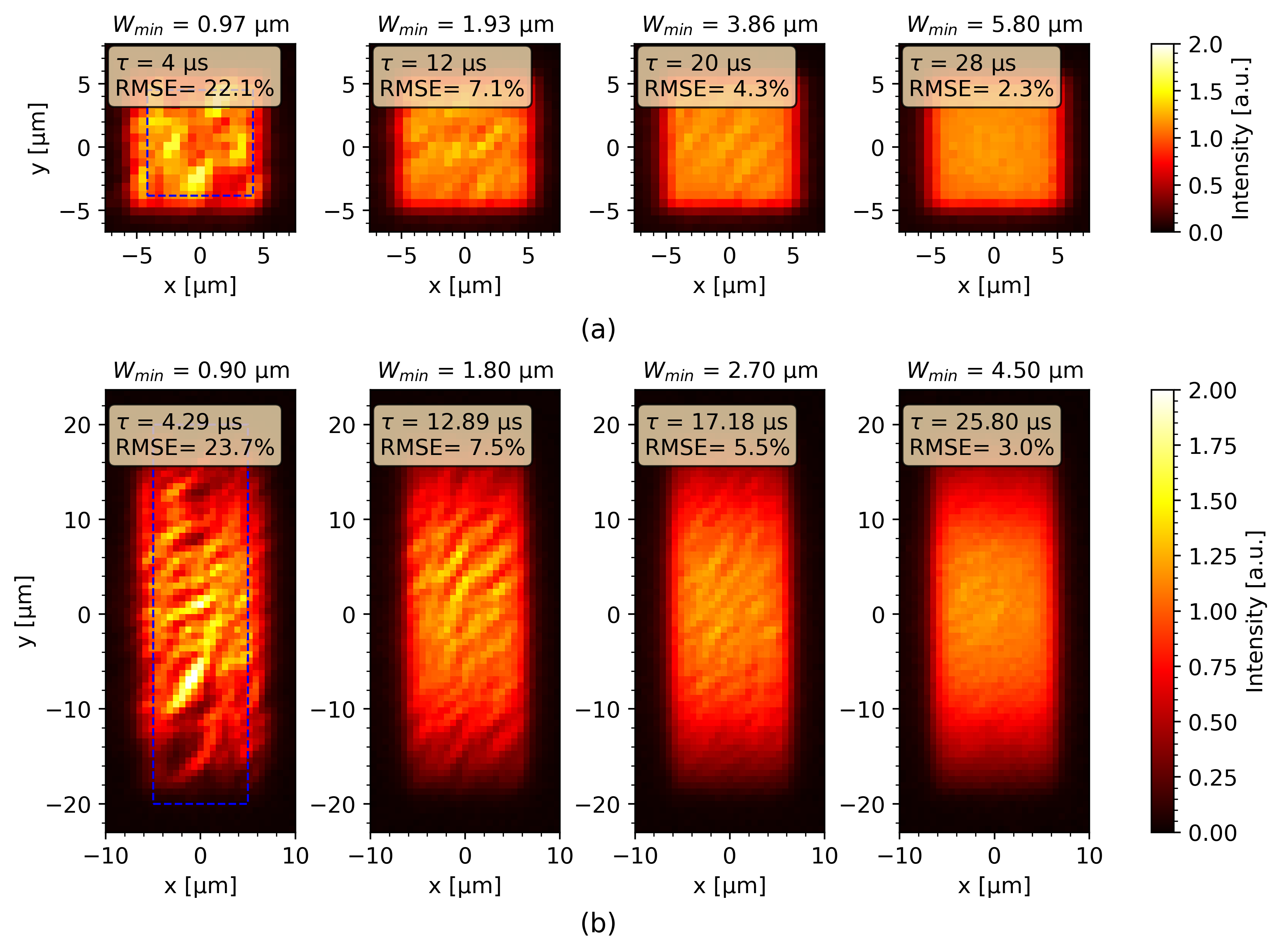}
    \caption{\textbf{Incommensurate staggered projections. } More accurate projections are obtained by increasing the staggering parameter $\Delta n_x$. This increases the minimal artifact distance $W_{min}$ but also increases the driving period $\tau$ according to Eq. \eqref{eq: Wmin} for $\Delta n_x\neq1$. For the regular case, $\Delta n_x=\Delta n_y=1$, we have $W_{min}=a=C/\tau$. For all images the RMSE is determined with respect to a coherent artifact free reference image, within the region of interest marked by blue dashed lines in the first column (see main text). (a) 8x8$\mu $m uniform square, projected with spot pitch $a=0.966$~$\mu$m, staggering parameters $\Delta n_x=1,3,5, 7$, (b) 10x40$\mu $m parabolic intensity pattern, projected with spot pitch $a=0.9~\mu$m, staggering parameters $\Delta n_x=1,3,4, 6$. }
        \label{fig: demonstration}
\end{figure}

This is illustrated in \autoref{fig: demonstration} for two separable example target patterns, a 8x8~$\mu$m uniform square: $I_{tar}(x,y)=I_0$ and a 10x40~$\mu$m rectangle with a parabolic intensity variation: $I_{tar}(x,y)=A-B y^2$. For the former case the relevant driving amplitudes were constant $A^x_k=A^x, A^y_p=A^y$, for the latter case they were determined via a least squares fit as described in the previous section. The phases $\alpha^j_k$ for both patterns were optimized via a Gerchberg-Saxton algorithm, see \autoref{appendix: gs}. 

As a measure of the projection accuracy we use a normalized root mean square error (RMSE), defined as the standard deviation of the relative residual intensity within the region of interest (ROI): \be \text{RMSE} = \sqrt{\frac{1}{N_{pxl}}\sum_{ROI}\left(\frac{I_\text{data} - I_\text{ref}}{I_\text{ref}}\right)^2 }\,, \label{RMSE}\ee where $N_{pxl}$ is the number of pixels within the ROI, and the residual is computed with respect to a reference image $I_{ref}$ that is free of coherent artifacts. These reference images were obtained experimentally, by using an incommensurate staggered lattice, with the same spot pitch, but  with $W_{min}>L$, where $L$ is the image width. Notice that the RMSE defined in this fashion does not measure the error induced by the PSF and finite spot pitch. It is solely dependent on the error induced by the coherent artifacts and on the shot noise from finite camera counts: $\textrm{RMSE}_{shot}\approx\sqrt{N_{pxl}^{-1}\sum_{ROI} N_{ref}^{-1} }$, with $N_{ref}$ the photon pixel count for the reference image. For the projected square of \autoref{fig: demonstration}a the estimated shot noise is $\textrm{RMSE}_{shot}=1.85\%$, for the parabolic pattern of \autoref{fig: demonstration}b we have $\textrm{RMSE}_{shot}=2.53\%$. Note that we obtain our images from projecting on a glass window (see \autoref{appendix: exp setup}) and work with low intensities to be well below the threshold for damaging the atom chip window. Note also that the camera exposure $t_{exp}$ is still orders of magnitude larger than the driving period, $t_{exp}\gg \tau$, such that the the acquired images indeed effectively show the time-averaged intensity. This is also the case for the other projection images in the paper. 

Finally note that in the case of sparse projections, for instance of separated optical tweezers, the resulting lattice of tweezers will be anisotropic ($\Delta n_x\neq \Delta n_y$). In case of the dense projections we consider in this work, the anisotropy does not introduce any limitation as this is compensated by the least squares fit to the target pattern. 

\newpage
\section{Comparison with line-scanning} \label{sec: comparison}

\begin{figure}[t]
     \centering
     \begin{subfigure}[l]{0.48\textwidth}
         \centering
         \includegraphics[width=\textwidth]{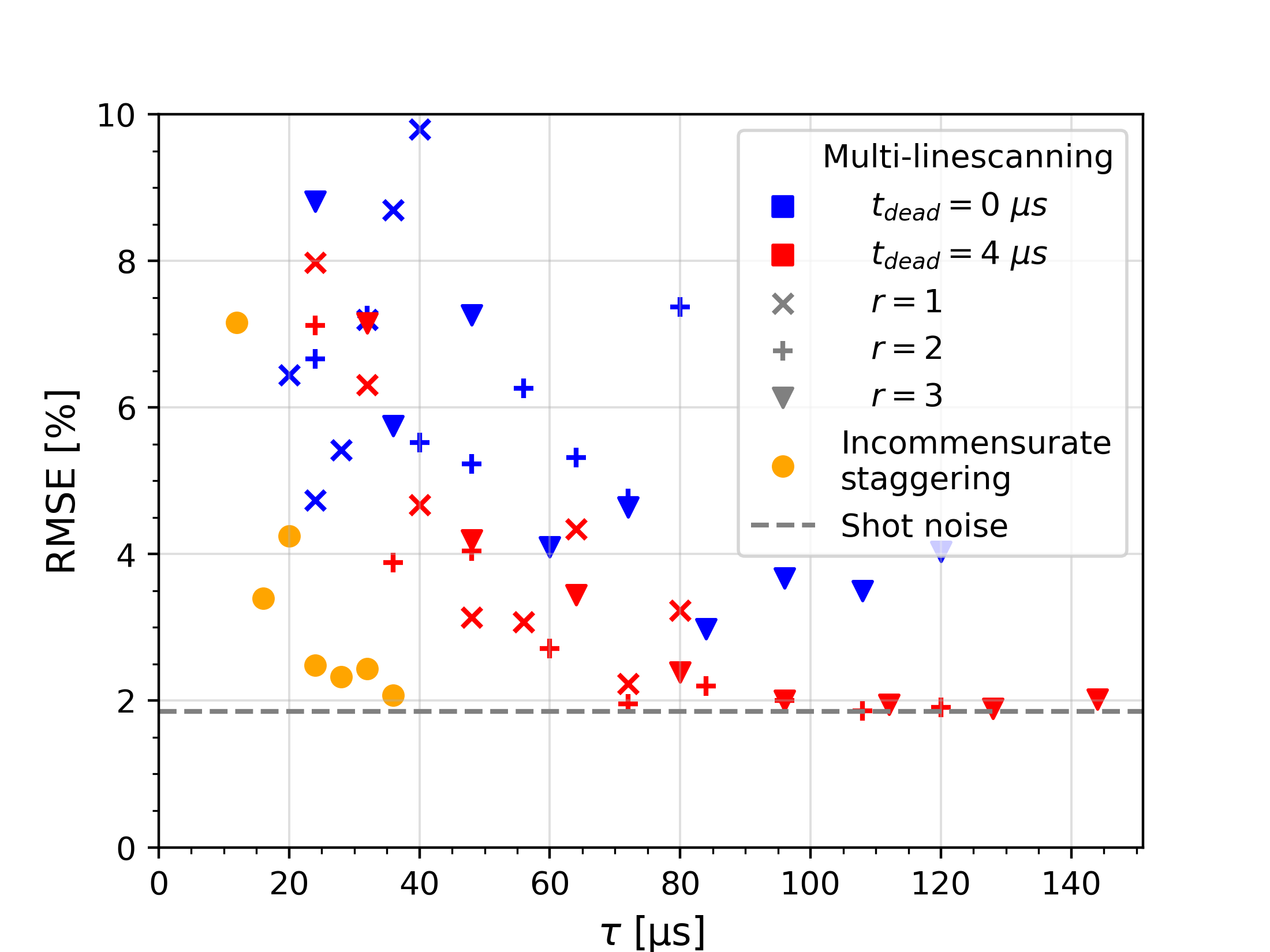}
                  \label{fig: speckle square}
                  \caption{}
     \end{subfigure}
    \begin{subfigure}[r]{0.48\textwidth}
         \centering
         \includegraphics[width=\textwidth]{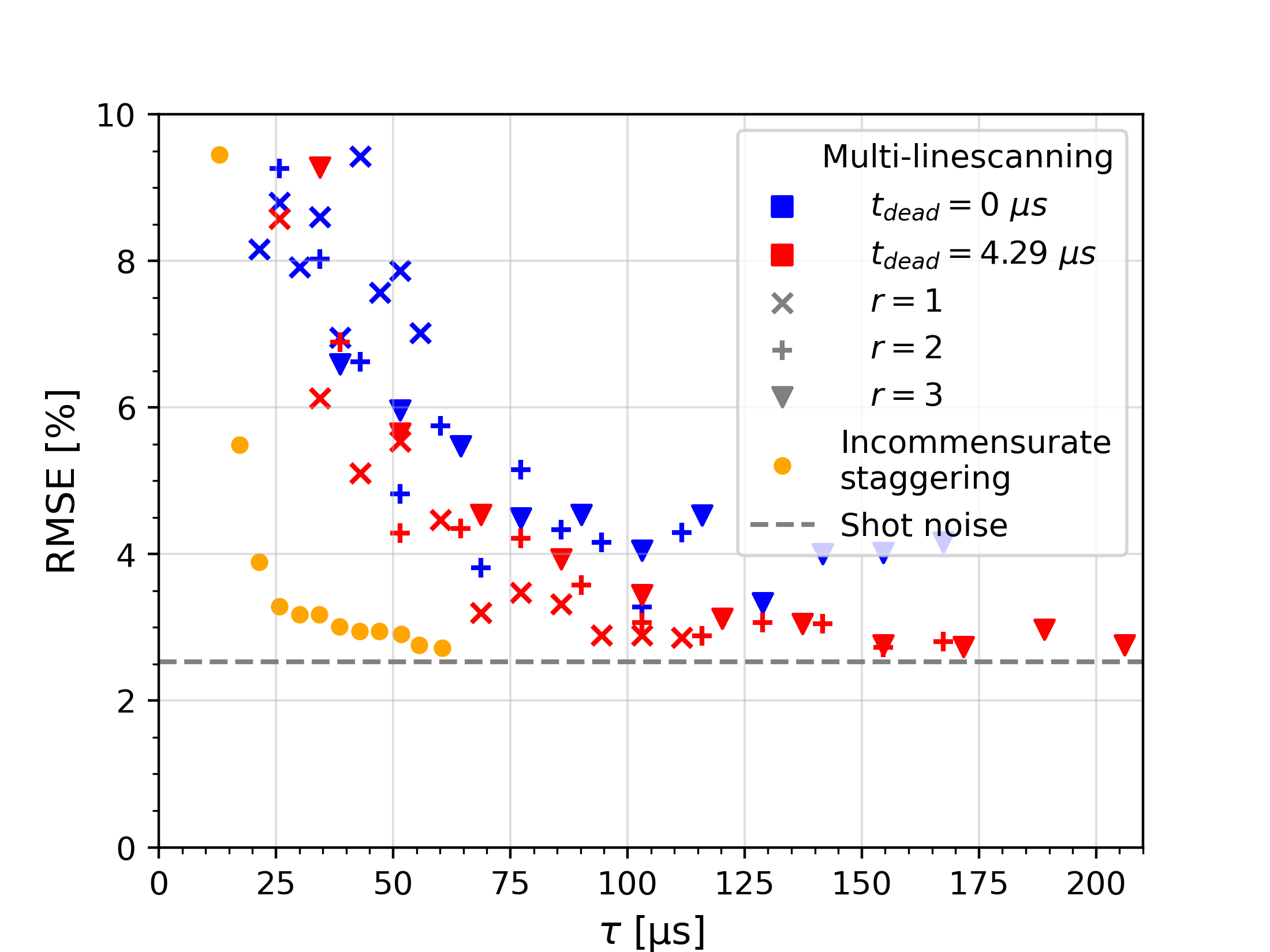}
                  \label{fig: speckle inv parab}
                  \caption{}
     \end{subfigure}         
        \caption{\textbf{Accuracy versus driving period for incommensurate staggering and multi-line scanning}. The root mean square error \eqref{RMSE} as a function of the total driving period $\tau$, for different AOD drives. Grey dashed lines show the estimated shot noise $\textrm{RMSE}_{shot}$. (a) Results for the square pattern of \autoref{fig: demonstration}a. The incommensurate staggered projections employed staggering parameters $\Delta n_x=1-8$, the multi-line scanning projections used interline distances (in spot pitch units) $N=1-10$, for different choices of the number of repetitions $r$ and the dead time $t_{dead}$ as indicated in the figure (see main text). All projections used the same spot pitch $a=0.966~\mu$m. (b): Results for the parabolic intensity pattern of \autoref{fig: demonstration}b. Here the staggering parameters are $\Delta n_x=1-14$, the employed interline distances are $N=1-13$ and the spot pitch is $a=0.9~\mu$m. Both figures only show the results that have a RMSE below 10\%.}
        \label{fig: comparison}
\end{figure}

As mentioned in the introduction, another technique to avoid coherent artifacts subdivides the image into a series of lines which are projected sequentially \cite{Treptow2021}. Taking the lines parallel to the $y$-axis, this entails sequential 1D multi-tone driving of the $y$-axis AOD, as in \autoref{appendix: 1D}, accompanied by synchronous sequential single-tone driving of the $x$-axis AOD. A general issue with such scanning protocols is the crosstalk noise \cite{Treptow2021}. During the brief simultaneous illumination of both waveforms inside the AOD, the projection is distorted. This is mitigated by repeating each frame $r$ times and by introducing a dead time $t_{dead}$ — during which $s(t)=0$ — in between frames \cite{Treptow2021,TreptowPhD}.

To have a fair comparison with line-scanning for the separable patterns of the previous section, we allow for multiple lines within one frame, as in \cite{TreptowPhD}, with neighboring lines in a single frame separated by a distance $\ell$ which is taken to be a multiple of the spot pitch $\ell=N\times a$, $N\in \mathbb{N}^+$. This is achieved by driving the $x$-axis AOD with a signal composed of the corresponding equidistant frequencies, with $\Delta f=N\times\frac{a}{C}$. The distance between neighboring lines $\ell$ in the image plane is then also the smallest coordinate distance between interfering spots on a $\Delta x=-\Delta y$ diagonal, $W_{min}=\ell$. \newpage

This results in a total projection period $\tau_{ls}$ for an image composed of the $N=W_{min}/a$ different frames: \be \tau_{ls}=\frac{W_{min}}{a}\left(\frac{C}{a} r+t_{dead}\right)\,.\ee Note that in the absence of crosstalk noise, one could take $r=1$, $t_{dead}=0$, which would actually lead to a smaller projection period than in the case of incommensurate staggering, \be \tau_{is}=\frac{C}{a}\left(1+\frac{W_{min}}{a}\right)\,, \ee from Eq. \eqref{eq: Wmin}. But, as our results in \autoref{fig: comparison} show, the image degradation due to crosstalk noise is quite considerable. To achieve a good accuracy one has to include a nonzero dead time in combination with repetitions, leading to larger driving periods. Our best multi-line scanning results were obtained by taking the dead time equal to the AOD period for the projection of a single frame, $t_{dead}=C/a$. We found smaller dead times (not shown in the figure) to yield worse results than a zero dead time. In any case, our results in \autoref{fig: comparison} show that the incommensurate staggering scheme clearly outperforms line-scanning strategies, it is always faster for a given accuracy.

\section{2D - non-separable patterns} \label{section: non sep}
As already commented, because of the separable control on the amplitudes of the different spots, see Eq. \eqref{eq: IC2D}, a general non-separable target pattern can not be well approximated by a single incommensurately staggered AOD projection. In that case a scanning strategy is required, where the full non-separable pattern is recovered as the time-average of a sequential projection of a number of separable patterns. While this requires a longer total projection time, we show here that our protocol still leads to a reasonably fast total projection of non-separable patterns. 

To approximate a general image as a series of separable sub-images we use the non-negative singular value decomposition (NNSVD) of \cite{WeixiangLiu2008}, see also \cite{TreptowPhD}. The target pattern $I_{tar}(x,y)$ is approximated as a sum over $\mathcal{R}$ separable sub-images: 
\be I_{tar}(x,y)\approx \sum_n^\mathcal{R} U_{n}(x)S_nV_{n}(y)\,, \ee
with $U_n(x),S_n,V_n(y)\geq0$. This amounts to an approximate positive matrix factorization, $I_{tar}\approx U S V^T$, with rank $\mathcal{R}$. Note that for our numerical scheme we work with discretized spatial variables, and $(x,y)$ therefore are countable indices. The approximation is found by minimizing the Frobenius norm of the residual $||I_{tar}-USV^T||^2$ with gradient descent \cite{WeixiangLiu2008}. As in \cite{TreptowPhD}, we take the element-wise absolute values of the regular SVD decomposition for the initial guess. Different from \cite{WeixiangLiu2008, TreptowPhD} we set a certain finite rank right from the start, which ensures that the obtained approximation is optimal for the given rank.

 Starting from a NNSVD approximation with rank $\mathcal{R}$, we determine the optimal driving amplitudes for each individual separable sub-image $I_n(x,y)=U_n(x)S_nV_n(y)$, as in the previous section. The full image is then obtained by scanning over the $\mathcal{R}$ sub-images, without repetitions, nor dead time. Note that the AOD transition time $(4~\mu s)$ is much smaller than the typical time to project a sub-image $(25~\mu s)$ which in itself reduces crosstalk noise. 

\begin{figure}[t]
     \centering
     \begin{subfigure}[b]{0.20\textwidth}
         \centering
         \includegraphics[width=\textwidth]{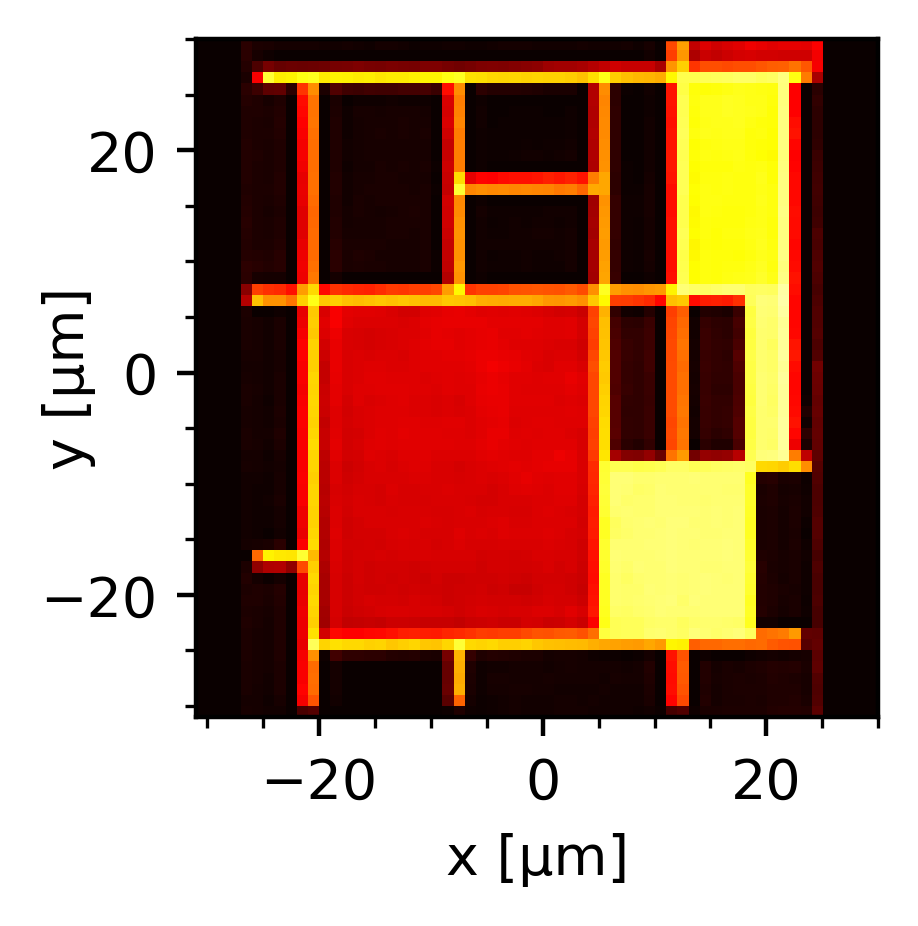}
        \caption{}
        \label{fig: mondriaan target hot}
     \end{subfigure}
     \hfill
     \begin{subfigure}[b]{0.78\textwidth}
         \centering
         \includegraphics[width=\textwidth]{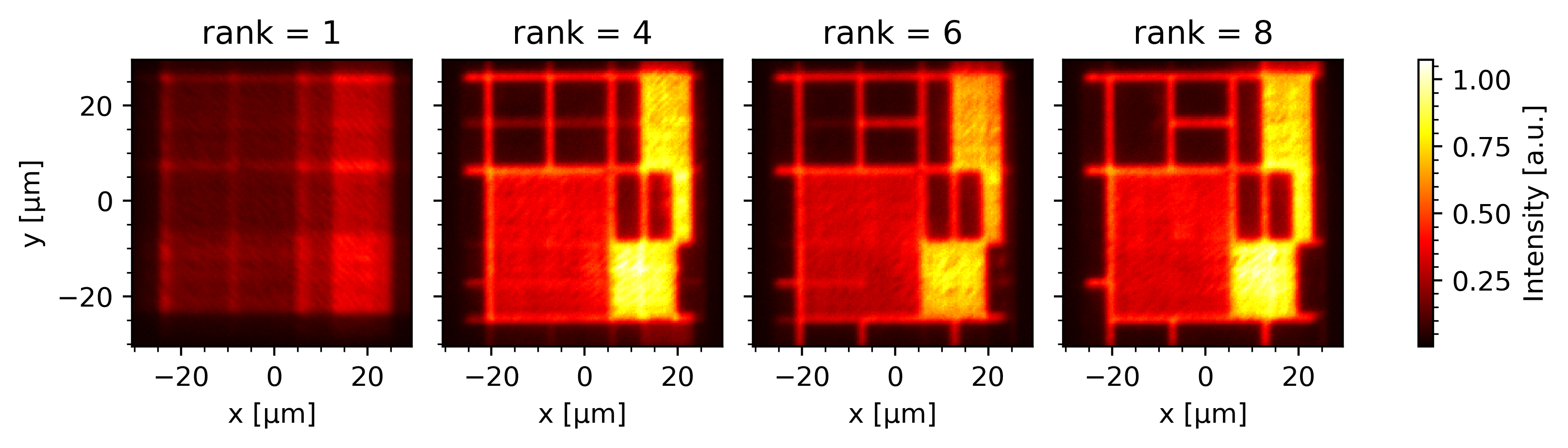}
         \caption{}
         \label{fig: mondriaan rank}
     \end{subfigure}
        \caption{\textbf{Projection of a grayscale Mondriaan.} (a) The target intensity, a pixelated 61x61 monochrome version of Piet Mondriaan's \textit{Composition en rouge, jaune, bleu et noir}. (b) Scanned projections for different rank approximations (see main text). All sub-images used a spot pitch $a=\!1~\mu m$ and staggering parameter $\Delta n_x=7$, implying a minimal interfering spot distance $W_{min}=6~\mu m$ and a sub-image projection time $t=27.05~\mu$s. We used neither repetitions, nor dead times between the sub-images. Total projection times $\tau=27.05,\, 108.2,\, 162.3,\, 216.4~\mu s$. }
    \label{fig: mondriaan}
\end{figure}

In \autoref{fig: mondriaan} we illustrate the result of this procedure on a non-separable example target pattern with sharp features and different grey scales. For such non-separable patterns we again get a clear trade-off between the image accuracy on the one hand, now determined by the spot pitch $a$, the minimal interfering spot distance $W_{min}$ and the approximation rank $\mathcal{R}$, and the total projection period $\tau$ on the other hand. Nevertheless, we find good image approximations at reasonable total projection periods. The final accuracy is also limited by the spot waist $w$ which explains why the quality of the fine lines in \autoref{fig: mondriaan} improve little when using more sub-images.

We stress that, as was the case for the projections of separable patterns, these projections were generated directly, without any feedback for further image improvement. Note also that our procedure is not optimal yet. 
We did not fully explore the inclusion of dead times. Furthermore, a better approach for determining the sub-image spot amplitudes may consist of a least square fit for all sub-images in one go; we leave this for future work. 
Finally, notice that a line-scanning scheme \cite{Treptow2021} would require the sequential projection of $N\approx L/a$, number of individual lines, with $L$ the image width in the scanning direction and $a$ the spot pitch (Eq. \ref{eq: spot pitch}). For the image in \autoref{fig: mondriaan}, we have $N=61$. This full projection would therefore take at least $N\times C/a=236~\mu$s. This is without any crosstalk mitigation. Based on \cite{Treptow2021} and results in \autoref{sec: comparison}, we can expect that a reasonable suppression of the crosstalk noise would at least double the total projection time.

\section{Discussion}\label{sec: conclusions}
In this work we have introduced a concrete scheme for fast and accurate AOD projections of 2D intensity patterns, based on appropriately chosen incommensurate lattices in the frequency domain. Our construction exposes a clear and quantitative trade-off between the multi-tone driving period of the two orthogonal AODs and the resulting suppression of coherent artifacts. As we have shown, for both separable and non-separable patterns, this trade-off is favorable when compared to other line-scanning strategies, enabling substantially faster projections at comparable accuracy. 

An important practical advantage of our method, in contrast to projections with DMDs or LC-SLMs, is that it does not rely on a feedback protocol. For separable target patterns, optimized AOD driving amplitudes can be obtained efficiently via a simple least-squares procedure, while extending the approach to non-separable patterns requires only an additional matrix factorization step with modest computational overhead. Furthermore, as discussed, our scheme can be easily augmented with a fast Gerchberg-Saxton algorithm to optimize the AOD driving phases for maximal power efficiency. Taken together, we expect that our method will prove useful for different types of applications that require the generation of arbitrary intensity patterns with fast and accurate performance.

Given the context of our AOD setup, see \autoref{appendix: exp setup}, a specific application that we have in mind, is the creation of arbitrary potentials for ultra-cold atoms, via the light-matter dipole interaction \cite{Amico2021}. While in this paper we have focused primarily on the time-averaged intensity pattern $I_{avg}(x, y)$, AODs are intrinsically time-dependent devices, giving rise to a temporal speckle $\tilde{I}(x,y,t)$ associated with the modulated part of the intensity pattern. Periodically driven AODs therefore effectively realize a form of Floquet dynamics, which can induce micromotion and heating \cite{Goldman2014}. We expect that the comparatively short driving periods that our scheme allows, will substantially mitigate such unwanted effects. A systematic theoretical and experimental investigation of this question, remains an important direction for future work, see the related results of \cite{Bell2018} and \cite{Sandberg2020}. Finally, a natural extension of the present work is to move beyond time-averaged static projections and explore the engineering of arbitrary time-dependent potentials. In this context, the driving period constitutes a fundamental limit for AOD-generated dynamic traps, and the short periods that our method allows may therefore offer a significant advantage, in addition to its feedback-free character.  

Another possible application is optogenetic stimulation, where light is used to trigger individual neurons in the brain. Our method gives access to sub-ms time resolution needed to resolve the fastest processes while maintaining the required accuracy for uniform triggering probabilities \cite{Ricci2022}. Note that here the required full 3D access could be obtained by combining our protocol with an optical setup as in  \cite{guo2025acoustoopticlens3dshuttling} for fast axial control.

\appendix

\section{1D multi-tone driving}\label{appendix: 1D}
To derive the relevant formula and to highlight the parallels and differences between 1D and 2D AOD multi-tone driving, we consider here the simpler case of 1D patterns generated by a single AOD. In our setup this is effectively realized by driving the $y$-axis AOD with a single tone and considering multi-tone driving for the $x$-axis AOD. \\

The transmission function for light passing through a driven AOD crystal with width $d$ is given by \cite{AOfund1981, Fourier_Goodman1969}
\be H(x', t) = e^{i \frac{2\pi d}{\lambda} n(\zeta)} = e^{i \frac{2\pi d}{\lambda} \xi s(\zeta)}\,.\ee
The refraction index $n(\zeta)$ is modulated by a driving signal $s(\zeta)$, where $\xi$ is a material-specific constant and $\zeta = x'/v + t$ includes the propagation speed $v$ of the sound wave in the crystal.
After the AOD and the relay optics, the microscope objective Fourier transforms the incoming light field $U_{in}(x')$ to the output light field $U_{out}(x,t)$ in the image plane as  \cite{Fourier_Goodman1969}:
\be U_{out}(x,t) = \mathcal{OFT}\left\{U_{in}(x')e^{i \frac{2\pi d}{\lambda} \xi s(\zeta)}\right\}\,, \ee
where the optical Fourier transform is given by 
\be\mathcal{OFT}\left\{u(x')\right\}=\dfrac{1}{i\lambda f_L}\int u(x') e^{-i\frac{2\pi}{\lambda f_L}x x'}dx',\ee
with $f_L$ the focal length of the objective. Parameterizing the driving signal $s(t)$ as a multi-toned signal with amplitudes $a_k$, frequencies $f_k$ and phases $\alpha_k$,
\be s(t) = \sum_k a_k \sin(2\pi f_k t + \alpha_k)\,,\label{eq:1Dmultitone}\ee
the expression for the transmission function can be rewritten via the Jacobi-Anger (JA) expansion:  
\begin{equation}
    H(x', t) 
    =e^{i\sum_kz a_k \sin\theta_k} = \prod_ke^{iz  a_k\sin\theta_k} \overset{\text{JA}}{=}\prod_k \sum_{q=-\infty}^{+\infty}J_q(z a_k)e^{iq\theta_k} \,,
\end{equation}
 with: 
\be \theta_k=2\pi f_k (x'/v + t)+\alpha_k, \qquad z =  2\pi d \xi / \lambda\,. \ee
The JA expansion is associated to the beam splitting into multiple diffraction orders, denoted by index $q$. Specifically, in the linear regime $z a_k\ll 1$, this results in the approximation: 
\begin{align}
    H(x', t)\approx \prod_k \left( 1 + \frac{za_k}{2}e^{i\theta_k} -\frac{za_k}{2}e^{-i\theta_k}\right) \approx 1+\sum_k\frac{za_k}{2}e^{i\theta_k} -\sum_k\frac{za_k}{2}e^{-i\theta_k},
\end{align}
yielding the undiffracted beam ($q=0$, first term) and the first diffraction orders ($q=\pm 1$, second and third term). Physically blocking the undiffracted beam and the $q=-1$ diffracted beam, we finally arrive at:

\begin{align}
    U_{out}(x, t) & \approx   \dfrac{z}{2i\lambda f_L} \sum_k  a_k e^{i2\pi f_k t + i\alpha_k} \int U_{in}(x') e^{-\frac{2\pi i}{\lambda f_L}\left[x-\frac{\lambda f_L}{v} f_k\right] x'}dx' \\
    &\cong \sum_k c_k a_k PSF(x- C f_k)e^{i2\pi f_k t + i\alpha_k} \\
    &=\sum_k A_k PSF(x- x_k)e^{i2\pi f_k t} \,.
\end{align}
Here, on the second line we drop an irrelevant factor $-i$, gather all (material) constants and abstract the incoming beam profile to the PSF. We also incorporate an additional frequency-dependence of the diffraction efficiency — not covered by the idealized derivation above — by introducing an explicit $k$-dependence for the total pre-factor $c_k$. On the final line we absorb this diffraction map, together with the AOD driving amplitude and phase into one complex number $A_k=c_k a_k e^{i\alpha_k}$ and denote $x_k=C f_k$ for the translation of the PSF in the image plane that corresponds to the frequency $f_k$. 

The actual projected light intensity $I(x, t)=U(x, t)\bar{U}(x, t)$, is then found as a double sum:    
\begin{align}
    I(x, t) = \sum_{k,l}I_{kl}(x, t)\,, 
    \end{align}
    with\begin{align}
    I_{kl}(x, t) =  A_k \bar{A_l} PSF(x-x_k)\overline{PSF}(x-x_l) e^{2\pi i (f_k - f_l)t} \,.\label{eq: 1D int}
\end{align}
This sum splits naturally into an \emph{incoherent} and \emph{coherent} part, where the incoherent terms ($k=l$) represent the spots corresponding to the individual tones in the driving signal. The cross terms ($k\neq l$) are called coherent because they arise due to interference between two different spots. 

\begin{figure}[t]
    \centering
   \includegraphics[width=0.8\linewidth]{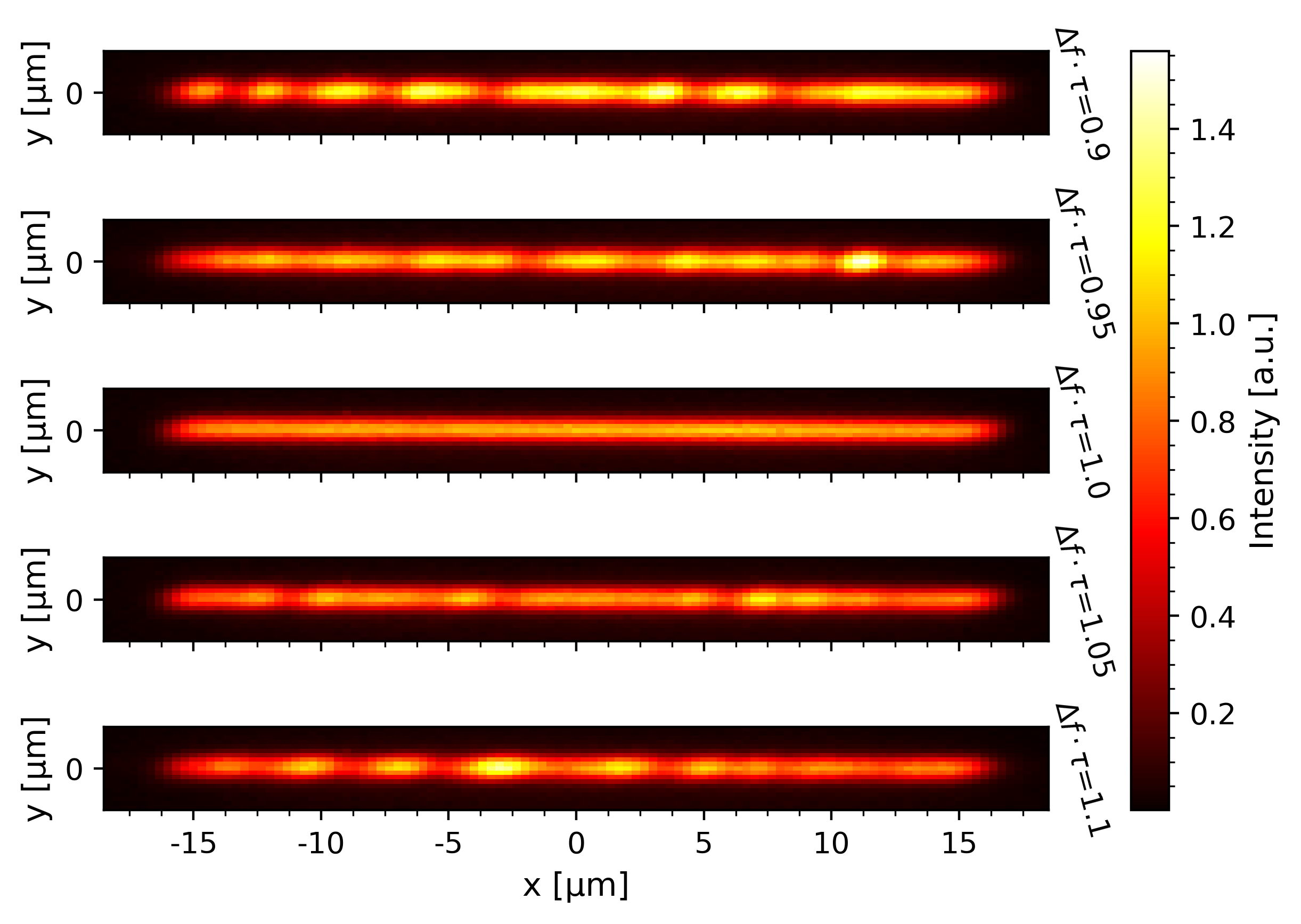}
    \caption{\textbf{Projected lines with different frequency steps.} For all the images the ($x$-axis) driving signal has the same period, $s(t)=s(t+\tau)$ with $\tau=10$~$\mu$s, and where for $\tau\in[0,10[\, \textrm{ms}$ the signal has the explicit form of Eq. \eqref{eq:1Dmultitone}, with $|A_k|=1$ and random choices for the phases $\alpha_k$. From the upper to the lower panel the signal frequencies read: $f_k=f_0+ k \Delta f$, with $f_0=f_c=525/\tau$, $\Delta f \cdot \tau =(0.9,0.95,1,1.05,1.1)$, and $k\in[-k_0/2,k_0/2]$, $k_0=(92,86,82,78,74)$.  }\label{fig: poster lines}
    \end{figure}

If we now restrict the driving signal to commensurate frequencies $f_k = n_k/\tau$, with $n_k\in \mathbb{N}$, or equivalently, if we work with a periodic signal $s(t)=s(t+\tau)$, the resulting light intensity also has a period $\tau$, with the coherent part $\tilde{I}(x, t)$ containing all components with nonzero frequencies $f=f_k-f_l=\frac{\Delta n}{\tau}\neq 0$ and the incoherent part containing the zero frequency components $I_{avg}(x)$:  
\begin{align}
    I(x,t) = I_{avg}(x) + \tilde{I}(x, t)\,,
\end{align}
with: 
\begin{align} \label{eq: I_avg 1d}
I_{avg}(x) =\frac{1}{\tau}\int^{t_0+\tau}_{t_0}\! dt \, I(x,t)=\sum_k I_{kk}(x, t)= \sum_k |A_k|^2|PSF(x-x_k)|^2\,.
\end{align}
This shows that in the case of a single AOD, periodic multi-tone driving yields an averaged image that is composed of a series of individual spots separated by the \emph{spot pitch} $a$, i.e. the distance between two neighboring spots:  $a = C \Delta f_{min} = C/\tau $.

As an example we show in the middle panel of \autoref{fig: poster lines}, a projected uniform line, $|A_k|=1$, employing frequencies $f_k=f_0+k \Delta f$ commensurate with the driving period, $f_0=n_0/\tau$ ($n_0\in \mathbb{N}$), $\Delta f=1/\tau$. Specifically, the projection used $\tau=10\,\mu\textrm{s}$ or corresponding spot pitch $a=C/\tau=0.39\,\mu\textrm{m}$, and a total of 83 subsequent spots (frequencies). The other panels illustrate unwanted artifacts that appear when we instead employ incommensurate frequencies, $ f'_k=f_0+k \Delta f'$, with $\Delta f'\neq 1/\tau$,  for the - in that case necessarily discontinuous - periodic driving signal $s(t)=s(t+\tau)$. In this latter incommensurate case, a certain frequency component $f'_k$, consists of several commensurate components $f_l$, which invalidates the simple one-to-one map between the frequency coefficients $A'_k$ and the actual spot amplitudes.

\section{Extension to 2D multi-tone driving}\label{appendix: 2D AOD formula}
Placing two AODs perpendicularly allows for sequential phase modulation of the light. Each AOD, driven by a separate function, diffracts its incoming light along one direction according to:
\be H(x', t) = e^{i \frac{2\pi d}{\lambda} \xi_x s_x(x'/v + t)} \,, \quad H(y', t) = e^{i \frac{2\pi d}{\lambda} \xi_y s_y(y'/v + t)} \, .\ee
The incoming light effectively sees a two-dimensional separable transmission function:
\begin{align}
    U_{out}(x,y,t) &= \mathcal{OFT}_{2D}\left\{U_{in}(x',y')H(x',y',t) \right\} \\
    &= \mathcal{OFT}_{2D}\left\{U_{in}(x',y')H(x', t)H(y', t) \right\}
\end{align}
Driving both AODs with multi-toned signals \eqref{eq:2Dmultitone}, the 1D derivation of the previous section can be applied to the separate transmission functions, remaining in the linear regime and singling out the first diffraction order in the JA expansion: 
\begin{align}
    U_{out}(x,y,t) =& \frac{1}{i\lambda f_L}\iint U_{in}(x',y')H(x',t)H(y',t)e^{-\frac{2\pi i}{\lambda f_L}[xx'+yy']}dx'dy'. \\
    \approx & \frac{z_x z_y}{4i\lambda f_L}\sum_{k,p} a^x_ka^y_p e^{i2\pi(f^x_k+f^y_p)t+i\alpha^x_k+i\alpha^y_p}\nonumber\\
    &\quad \iint U_{in}(x',y') e^{-\frac{2\pi i}{\lambda f_L}\left(\left[x-\frac{\lambda f_L}{v}f^x_k \right]x'+\left[y-\frac{\lambda f_L}{v}f^y_p \right]y'\right)}dx'dy' \\
    & \cong\sum_{k,p}c^x_kc^y_pa^x_ka^y_p ~PSF(x-Cf^x_k, y-Cf^y_p)e^{i2\pi(f^x_k+f^y_p)t+i\alpha^x_k+i\alpha^y_p}\,,
\end{align}
where we now incorporate the additional frequency dependence of the diffraction efficiency for each AOD - again not covered by the idealized derivation - in the separable $(k,p)$-dependent pre-factor $c^x_k c^y_p$. This latter equation serves as the starting point for the results in the main text.

\section{Gerchberg-Saxton algorithm for phase optimization} \label{appendix: gs}
The averaged image $I_{avg}(x)$ in the 1D case, Eq. \eqref{eq: I_avg 1d}, and the incoherent part of the averaged image $I_{IC}(x,y)$ in the 2D case, Eq. \eqref{eq: IC2D}, do not depend on the phases in the driving signal. These can therefore be chosen freely, without altering the averaged images, to maximally reduce amplitude variations in the driving signals. For this we implemented a variation on the Gerchberg-Saxton algorithm, see also \cite{yang2015}. 

\begin{figure}[t]
  \centering
  \includegraphics[width=\textwidth]{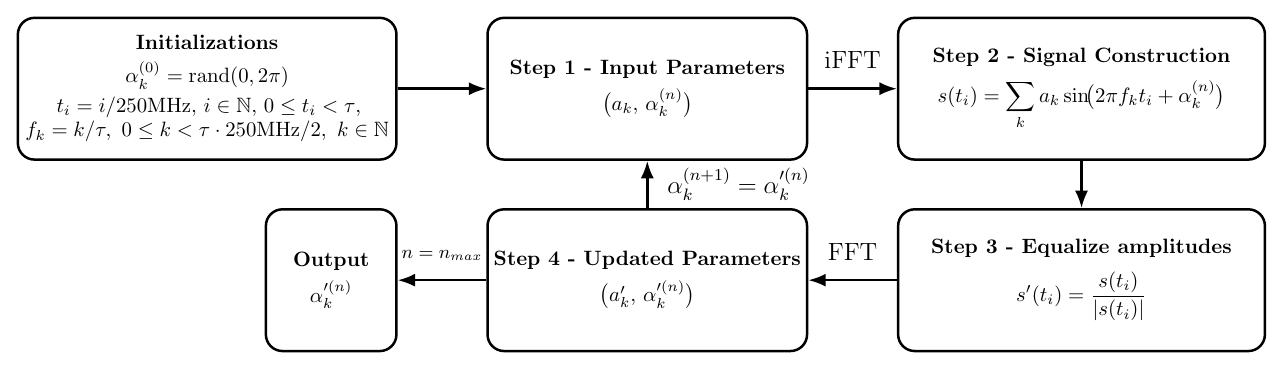}
  \caption{\textbf{Gerchberg-Saxton algorithm.} Illustration of waveform amplitude reduction scheme; see main text.}
  \label{fig:signal-workflow}
\end{figure}

As illustrated in \autoref{fig:signal-workflow}, we consider a periodic driving signal $s(t)$ of the form \eqref{eq:1Dmultitone} with frequencies $f_k=n_k/\tau$, amplitudes $a_k$, phases $\alpha_k$ and a period $\tau$ commensurate with the AWG sample rate: $N_\text{samples}=\tau \cdot 250\text{MHz} \in \mathbb{N}$, see \autoref{appendix: exp setup}. The discretized driving signal results from sampling Eq. \eqref{eq:1Dmultitone} at the timestamps $t_i=i/250\text{MHz}$. A step $n$ in the iterative scheme starts from the given amplitudes $a_k$ and certain phases $\alpha_k^{(n)}$. The amplitudes in the resulting driving signal are then renormalized to one: $s'(t_i) = s(t_i)/|s(t_i)|$. Numerous alternatives exist to this step \cite{yang2015}, but no relevant improvement was found over the presented scheme. This signal is transformed back to the frequency domain, resulting in coefficients $a'_k e^{i{\alpha'_k}^{(n)}}$. The next iteration then starts again from the original amplitudes $a_k$, but with updated phases $\alpha_k^{(n+1)}={\alpha'_k}^{(n)}$. After $n_{max}$ iterations, the procedure halts. We typically find near convergence of the reduction in signal amplitude for $n_{max}=100$. In the case of 2D projections, this procedure is applied separately to both $s_x(t_i)$ and $s_y(t_i)$. 

\section{Coherent artifact measurement}\label{appendix: coherentartifact} 

\begin{figure}[H]
    \centering
    \includegraphics[width=\textwidth]{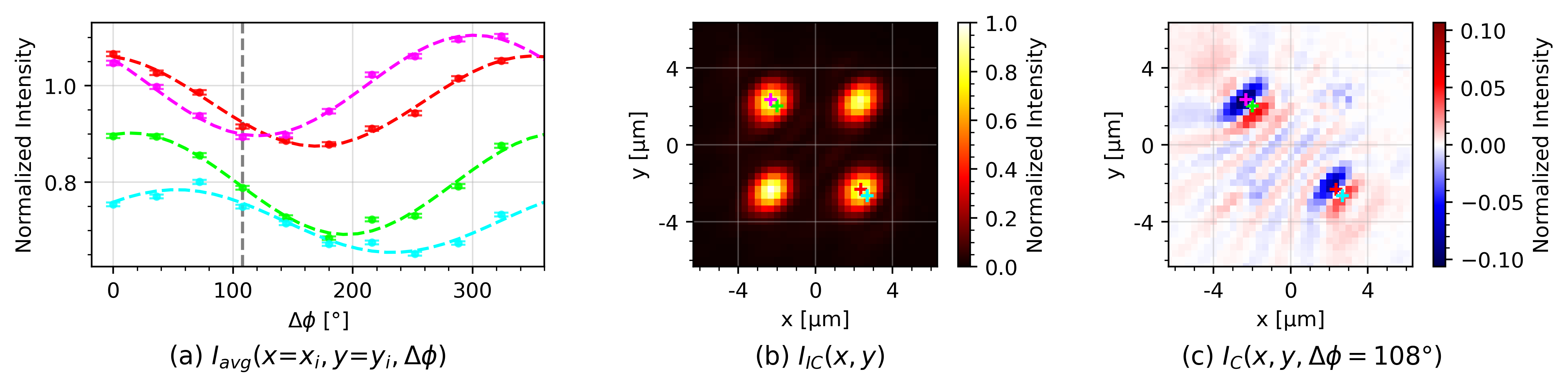}
    \caption{\textbf{Examples of coherent artifact measurements. }
    The coherent artifacts are measured by imaging $I_{avg}(x,y;\Delta\phi)$, see Eq. \eqref{eq: CAmeasurement}, for different values of $\Delta\phi$. At each location $(x_i, y_i)$, a cosine is fitted. The data shown here has an inter-spot distance $W=4.63~\mu$m ($\Delta f=1.2$MHz). 
    (a) Examples of cosine fit for four marked locations ('+' in (b) and (c)). The data points show the measured intensity $I_{avg}(x_i, y_i, \Delta\phi)$ where the error is estimated from camera shot noise ($0.5\%$). Colored dashed lines show the fitted cosine from which the artifact amplitude, artifact phase and incoherent intensity are extracted. 
    The gray dashed line marks the $\Delta \phi$ of (c). 
    (b) The incoherent part of the intensity $I_{IC}(x,y)$. 
    (c) Direct image of the coherent artifact $I_{C}(x,y; \Delta\phi=\!108\degree)=I_{avg}(x,y; \Delta\phi=\!108\degree) - I_{IC}(x, y)$.}
    \label{fig: 2spotInterference excerpts cos}
\end{figure}


\begin{backmatter}
\bmsection{Funding}
This work is supported by FWO grants I004520N and  G0C5123N. CT also acknowledges support by FWO grant 1176525N.
\bmsection{Acknowledgment}
We thank Ra\'{u}l Bola, Frederic Català i Castro, Ignacio Aquilino Martínez Sánchez, Seth Caliga, Kasper Van Gasse and Filip Beunis for interesting discussions and we thank Frank Verstraete for his support in setting up the lab. 
\bmsection{Disclosures}
The authors declare no conflicts of interest.
\bmsection{Data Availability Statement}
Data underlying the results presented in this paper are available in Ref.\cite{data_zenodo}.
\end{backmatter}
\bibliography{2DAOD}

\end{document}